\documentclass[epj,numbook]{svjour}
\usepackage[USenglish]{babel}
\usepackage{amsmath}
\usepackage{amssymb}
\usepackage{bbm} %matrix identity with \mathbbm{1}
\usepackage{graphicx}
\usepackage[pdftex,pdfauthor={Thomas Vonk, Ulf-G. Mei{\ss}ner},pdftitle={The Baryon-Baryon Interaction in the Large-Nc Limit},hidelinks,colorlinks=true,allcolors=blue]{hyperref}
\usepackage{cite}
\usepackage[dvipsnames]{xcolor}

\newcommand{\tr}[1]{\left\langle #1 \right\rangle}
\newcommand{\com}[2]{\left[ #1,#2 \right]}
\newcommand{\acom}[2]{\left\{ #1,#2 \right\}}
\newcommand{\sdot}[2]{\left( #1\cdot #2 \right)}
\newcommand{\order}[1]{\mathcal{O}\left( #1 \right)}
\newcommand{\titlemath}[1]{\texorpdfstring{{\boldmath $#1$}}{}}
\newcommand{\B}{\mathcal{B}}
\newcommand{\football}{\scalebox{0.66}{\rotatebox[origin=c]{90}{O}}}
\renewcommand{\Im}{\operatorname{Im}}

\makeatletter
\newcommand{\Biggg}{\bBigg@{4}}
\makeatother

\begin{document}
\title{The Baryon-Baryon Interaction in the Large-{\boldmath$N_c$} Limit}

\author{Thomas Vonk\inst{1}\thanks{e-mail: vonk@hiskp.uni-bonn.de} \and Ulf-G. Mei{\ss}ner\inst{1,2,3}\thanks{e-mail: meissner@hiskp.uni-bonn.de}}
\institute{Helmholtz-Institut für Strahlen- und Kernphysik and Bethe Center for Theoretical Physics, Universität Bonn, D-53115 Bonn, Germany\and Institute~for~Advanced~Simulation~(IAS-4), Forschungszentrum~J\"{u}lich, 
D-52425~J\"{u}lich,~Germany\and Peng Huanwu Collaborative Center for Research and Education, International Institute for Interdisciplinary and Frontiers,
Beihang University, Beijing 100191, China} 

\date{Received: date / Revised version: date}

\abstract{
We analyze the large-$N_c$ structure of the baryon-baryon potential derived in the framework of SU(3) chiral perturbation
theory up to next-to-leading order including contact interactions as well as one-meson and two-meson exchange diagrams.
Moreover, we assess the impact of SU(3) symmetry breaking from a large-$N_c$ perspective and show that the leading order
results can successfully be applied to the hyperon-nucleon potential. {Our results include a reduction of the number of relevant low-energy constants of the leading order contact interaction from fifteen to three, and we show that consistency is preserved if the $F/D$ ratio is given by $2/3$ and the $C/D$ ratio for the baryon decuplet-to-octet coupling is given by 2.}
%\PACS{
%	{}{}   \and
%	{}{} 
%}
}

\maketitle

\section{Introduction}
While ordinary matter is largely made of the light up and down quarks, strangeness offers
a new dimension in the formation of matter and the possible forms of exotic matter, see the reviews \cite{Schaffner-Bielich:2008zws,Hiyama:2009zz,Gal:2016boi,Tolos:2020aln}.
{One manifestation of this} additional degree of freedom are the so-called hypernuclei, where one
or two hyperons are bound together with neutrons and protons. {These systems often feature}
unusual properties, e.g. the hypertriton, a bound state of a proton, a neutron and a
$\Lambda$ hyperon exhibits a matter radius of about 10~fm, which is gigantic on nuclear
scales, see e.g.~\cite{Hildenbrand:2019sgp}. To understand such types of systems, a
precise knowledge of the underlying baryon-baryon interactions
is required. This, however, is a formidable task as very few scattering data and a limited
number of hypernuclei are known. Another intriguing aspect is the appearance of hyperons
in dense neutron matter, which naively leads to a softening of the equation of state so that
neutron stars with 2 solar masses can not be sustained, though we know that these
exist~\cite{Demorest:2010bx,Antoniadis:2013pzd}.
This apparent ``hyperon puzzle'' can be solved with repulsive three-baryon forces or more
exotic mechanisms, but again it requires an accurate understanding of the interaction between
baryons to really understand such forms of matter,
see e.g.~\cite{Schulze:2011zza,Lonardoni:2014bwa,Tong:2024egi} and references therein.
In addition, comparing the baryon-baryon
interactions with the well studied and precisely understood nucleon-nucleon interactions
tells us about the breaking of the SU(3) flavor symmetry, which is generated by the
very different mass scales of the strange quark and the light quark masses. Therefore,
given the scarcity of experimental data on baryon-baryon and multi-baryon interactions,
theoretical approaches that go beyond the flavor SU(3) are very much welcome to help
guide the research in strange matter formation and {the understanding of the properties} of such
intriguing systems.

A quite worthwhile approach is the large-$N_c$ limit sche\-me introduced by 't Hooft \cite{tHooft:1973alw} as a means
of studying QCD amplitudes in a systematic way using the number of colors $N_c$ as an ordering parameter.
This was taken up by Witten \cite{Witten:1979kh}, who demonstrated the beneficial application of this scheme to
hadrons introducing a Hartree-like picture of large-$N_c$ baryons and establishing major results which also
lie at the basis of the present work, see Section~\ref{sec:largeNCBB}. Shortly after this, not only
the connection to the Skyrme model \cite{Skyrme:1961vq,Skyrme:1962vh} could be uncovered \cite{Witten:1983tx,Gervais:1984rc},
but also the fact that baryons with an $\text{SU}(N_f)\times\text{SU}(2)_\text{spin}$ symmetry come with an exact
contracted $\mathrm{SU}(2N_f)$ spin-flavor symmetry in the large-$N_c$ limit leading to a tower of degenerate
$\mathrm{SU}(N_f)$ baryon multiplets \cite{Gervais:1983wq,Dashen:1993as,Dashen:1994qi}. This allowed for a systematic
expansion of the Hartree Hamiltonian in terms of an SU(2$N_f$) operator basis \cite{Carone:1993dz,Luty:1993fu,Dashen:1993jt}.
The subsequent years saw successfull applications to the study of large-$N_c$ baryon masses
\cite{Jenkins:1993zu,Jenkins:1995gc,Jenkins:1996de,Oh:1999yj,Flores-Mendieta:2024mxh}, the nucleon-nucleon system
\cite{Kaplan:1995yg,Kaplan:1996rk,Banerjee:2001js,Phillips:2014kna,Samart:2016ufg,Richardson:2022hyj}, meson-baryon scattering \cite{Lutz:2001yb,Flores-Mendieta:2014vaa,Heo:2019cqo,Flores-Mendieta:2021wzh,Heo:2022huf}, {and three-nucleon
forces~\cite{Phillips:2013rsa,Epelbaum:2014sea}}.
Furthermore, the SU(3) baryon-baryon interaction has been studied in this framework {focussing on leading order chiral contact interactions \cite{Liu:2017otd}. The main goal of the present paper is hence to extend this previous work and to give an overall survey of all relevant contributions up to next-to-leading order in chiral power counting including one- and two-meson-exchange contributions.}

In the following sections we will hence analyze all ingredients of the baryon-baryon potential up to
next-to-leading order in SU(3) chiral perturbation theory from a large-$N_c$ perspective, that is leading
and next-to-leading order contact interactions in Section~\ref{sec:bbchiralcontact}, and one-meson and
two-meson exchange contributions in Sections~\ref{sec:bbchiralOME} and~\ref{sec:bbchiralTME}, respectively.
This will of course require an adequate introduction into the baryon-baryon interaction in the large-$N_c$ limit
which directly follows this introduction in the next section, where we will derive and analyze the general
structure of the large-$N_c$ baryon-baryon potential.

\section{Large-\titlemath{N_c} baryon-baryon interaction}\label{sec:largeNCBB}
\subsection{Contracted SU(6) spin-flavor symmetry and Hamiltonian}
It is well known that the baryon sector of QCD in the large-$N_c$ limit has an exact SU($2N_f$) spin-flavor symmetry \cite{Gervais:1983wq,Dashen:1993as,Dashen:1994qi} and that large-$N_c$ baryons can be described by a Hartree-like approximation \cite{Witten:1979kh}. The Hartree Hamiltionian for $N_f=3$ baryons can be constructed in terms of the operators {
\begin{align}
\hat{\mathcal{S}}^i & = q^\dagger\left(\frac{\sigma^i}{2} \otimes \mathbbm{1} \right) q,\nonumber\\
\hat{\mathcal{T}}^a & = q^\dagger\left(\mathbbm{1} \otimes \frac{\lambda^a}{2} \right) q,\label{eq:HartreeGenerators}\\
\hat{\mathcal{G}}^{ai} & = q^\dagger\left(\frac{\sigma^i}{2} \otimes \frac{\lambda^a}{2} \right)q\nonumber
,
\end{align}
which are the generators of the contracted SU(6) spin-flavor symmetry.} Here, $q = (u,d,s)$ represents a three flavor bosonic quark operator that carries no color, the $\sigma_i$'s are the three Pauli spin matrices and the $\lambda_a$'s are the eight Gell-Mann matrices. {The commutation relations of the corresponding Lie Algebra are given in Appendix \ref{app:su6commu}.} In this basis, the Hartree Hamiltonian is given by \cite{Carone:1993dz,Luty:1993fu,Dashen:1993jt,Kaplan:1996rk}
\begin{equation}\label{eq:HartreeHamil}
\hat{\mathcal{H}} = N_c \sum_n \sum_{s,t,u} h_{stu}
\left(\frac{\hat{\mathcal{S}}}{N_c}\right)^s\left(\frac{\hat{\mathcal{T}}}{N_c}\right)^t
\left(\frac{\hat{\mathcal{G}}}{N_c}\right)^u \delta_{s+t+u,n}\ ,
\end{equation}
where the coefficients $h_{stu}$ are of $\order{1}$ in the large-$N_c$
power counting. As this Hamiltonian must be rotation and SU(3) flavor invariant, the vector,
spin, and flavor indices suppressed in Eq.\,\eqref{eq:HartreeHamil} are fully contracted with each
other meaning that the coefficients $h_{stu}$ are tensors of any rank necessary to combine with the
respective generators from Eq.\,\eqref{eq:HartreeGenerators} to form rotational invariant objects.

The spin-flavor generators are supposed to act on bar\-yon states, which in the large-$N_c$ limit
consist of $N_c$ quarks and are totally symmetric in spin-flavor Fock space. In order to get
reasonable large-$N_c$ equivalents of the real-world baryons with half-integer spins, $N_c$ needs
to be odd.

The contracted SU($2N_f$) spin-flavor symmetry satisfied by $\hat{\mathcal{H}}$ leads to a tower
of SU($N_f$) baryon multiplets \cite{Gervais:1983wq,Dashen:1994qi}. For $N_f=3$, we adopt the
common approach and set the large-$N_c$ equivalent of the $N_c=3$ flavor octet baryons as being
those with spin $S=\tfrac{1}{2}$, and isospin and strangeness of $\order{1}$.
\subsection{Sources of large-\titlemath{N_c} suppression}
In order to distinguish large-$N_c$ baryon states $\B$ from ordinary baryons at $N_c=3$,
we use the curved bra-ket notation \cite{Luty:1993fu}. For the large-$N_c$ scalings of the
matrix elements between such baryon states $|\B)$ and $|\B^\prime)$ one finds
for the generators of Eq.\,\eqref{eq:HartreeGenerators}
\begin{equation}
(\B^\prime|\hat{\mathcal{S}}^i |\B) \sim 1,
\end{equation}
and \cite{Dashen:1994qi}
\begin{align}
(\B^\prime| \hat{\mathcal{T}}^a |\B) & \sim 1, & (\B^\prime|
\hat{\mathcal{G}}^{ai} |\B) & \sim N_c, & \text{for } a
& = 1,2,3, \nonumber\\
(\B^\prime| \hat{\mathcal{T}}^a |\B) & \sim \sqrt{N_c},
& (\B^\prime| \hat{\mathcal{G}}^{ai} |\B) & \sim \sqrt{N_c},
& \text{for } a & = 4,5,6,7,\nonumber\\
(\B^\prime| \hat{\mathcal{T}}^a |\B) & \sim N_c,
& (\B^\prime| \hat{\mathcal{G}}^{ai} |\B) & \sim 1, & \text{for } a & = 8, \label{eq:TGscalings}
\end{align}
where the more differentiated large-$N_c$ scalings of the latter are valid only for baryons
with strangeness of $\order{1}$. {The origin of these asymmetric scalings can be best understood in the quark picture: if the interacting baryons have strangeness of $\order{1}$, there are only $\order{1}$ possibilities of picking up a strange quark but $\order{N_c}$ possibilities of finding an up or down quark. }

This set of large-$N_c$ scaling rules dictates already a
large part of the $1/N_c$ power counting of the baryon-baryon interaction to be discussed
in more detail below.

Another source of large-$N_c$ suppression stems from the general momentum structure of the
resulting potential. Considering the fact that the baryon masses $m_B$ scale $\sim N_c$ and
are degenerate up to corrections relatively suppressed by $1/N_c^2$, the only way of achieving
a consistent matching to any \textit{low}-energy theory is to assume that the baryon momenta
scale as $\order{N_c^0}$, in which case for the baryon velocity and non-relativistic
energy one has $|\mathbf{v}|\sim E\sim 1/N_c$ \cite{Banerjee:2001js}, which at the same time
justifies a static limit approach to the baryon-baryon potential. Let $\mathbf{p}$ and
$\mathbf{p}^\prime$ denote the initial and final center-of-mass momenta of the baryons, then
the momentum transfer $\mathbf{q}$ and the momentum sum $\mathbf{k}$ are given by
\begin{equation}\label{eq:momenta}
\mathbf{q}=\mathbf{p}^\prime-\mathbf{p},\qquad
\mathbf{k}=\mathbf{p}^\prime+\mathbf{p},
\end{equation}
which both are considered independent of $\order{1}$ in the large-$N_c$ power counting \cite{Kaplan:1996rk}. Moreover, the energy transfer $q_0 = E^\prime-E$ in the non-relativistic
limit is given by $q_0 = \Delta m_B + \sdot{\mathbf{k}}{\mathbf{q}}/(2m_B)$ with $\Delta m_B$ the baryon mass splitting. In sum, this leads to the following large-$N_c$ scalings of the  quantities that finally enter the baryon-baryon potential:
\begin{align}
m_B & \sim N_c, &  |\mathbf{q}|^2 & \sim 1, & q_0  & \sim N_c^{-1},\nonumber\\
\Delta m_B & \sim N_c^{-1}, &  |\mathbf{k}|^2 & \sim 1, & \sdot{\mathbf{k}}{\mathbf{q}} & { \sim 1}.\label{eq:momentascalings}
\end{align}
Moreover, expanding the
baryon-baryon potential in a Taylor series of the above momenta leads to the second source of
$1/N_c$ suppressions due to factors of $1/m_B$. As argued in \cite{Kaplan:1996rk}, this
suppression follows the general rule that terms proportional to $\mathbf{q}^m\mathbf{k}^n$
are suppressed by
\begin{equation}\label{eq:qksuppression}
1/N_c^{\operatorname{min}(m,n)} .
\end{equation}

\subsection{The resulting baryon-baryon potential}
The general form of the Hartree baryon-baryon potential is found by calculating the matrix elements
\begin{equation}
V_{\B^\alpha\B^\beta\to \B^\gamma\B^\delta} = \left( \mathbf{p}^\prime,\gamma; -\mathbf{p}^\prime, \delta\left|\hat{\mathcal{H}} \right| \mathbf{p},\alpha; -\mathbf{p}, \beta\right),
\end{equation}
where $\alpha,\dots,\delta$ denote internal quantum numbers such as spin or flavor. For the SU(3) flavor symmetry case, it has been derived in the appendix of Ref.~\cite{Kaplan:1996rk}. Here, we do not separate out terms involving explicit SU(3)
breaking and stay within the operator basis of full SU(6) spin-flavor symmetry,
Eq.~\eqref{eq:HartreeGenerators}. Sources of isospin and SU(3) breaking will nevertheless
be discussed in due course. Adopting the notation of
Ref.~\cite{Kaplan:1996rk}, $\hat{\Lambda}^M$ may denote any of the spin-flavor
generators of Eq.~\eqref{eq:HartreeGenerators} with proper normalization
$\hat{\mathcal{S}}^i/\sqrt{3}$, $\hat{\mathcal{T}}^a/\sqrt{2}$, and $\sqrt{2}\hat{\mathcal{G}}^{ia}$.
The expansion of Eq.~\eqref{eq:HartreeHamil} eliminating redundant terms then yields
\begin{align}
\begin{split}
\label{eq:HamilExpansion}
V & _{\B^\alpha\B^\beta\to \B^\gamma\B^\delta} = N_c \sum_{n=0}^{N_c} v_{0,n} \left( \frac{\hat{\Lambda}_1 \cdot \hat{\Lambda}_2}{N_c^2}\right)^n \\
%\\ & \phantom{+}  N_c \sum_{n=0}^{N_c} v_{0,n} \left( \frac{\hat{\Lambda}_1 \cdot \hat{\Lambda}_2}{N_c^2}\right)^n \\
& +  N_c \sum_{n=0}^{N_c-1} v_{1,n} (\mathbf{q} \times \mathbf{k})^i \left( \frac{\hat{\mathcal{S}}^i_1 + \hat{\mathcal{S}}_2^i}{\sqrt{3}N_c}\right) \left( \frac{\hat{\Lambda}_1 \cdot \hat{\Lambda}_2}{N_c^2}\right)^n \\
& + N_c \sum_{n=0}^{N_c-2} v_{2,n} (\mathbf{q} \times \mathbf{k})^i \left( \frac{\hat{\mathcal{G}}_2^{ia}
\hat{\mathcal{T}}^a_1 + \hat{\mathcal{G}}_1^{ia}\hat{\mathcal{T}}_2^a}{N_c^2}\right) \left(
\frac{\hat{\Lambda}_1 \cdot \hat{\Lambda}_2}{N_c^2}\right)^n\\
& +  N_c \sum_{n=0}^{N_c-3} v_{3,n} (\mathbf{q} \times \mathbf{k})^i \left(2 \frac{\hat{\mathcal{G}}_1^{ia}
\hat{\mathcal{G}}_2^{ja}\hat{\mathcal{S}}^j_1 + \hat{\mathcal{G}}_2^{ia}\hat{\mathcal{G}}_1^{ja}
\hat{\mathcal{S}}_2^j}{\sqrt{3}N_c^3}\right)
\\ & \qquad\qquad\qquad\times \left( \frac{\hat{\Lambda}_1 \cdot \hat{\Lambda}_2}{N_c^2}\right)^n \\
& +  N_c \sum_{n=0}^{N_c-2} \Biggl[ v_{4,n} \left( \mathbf{q}^i\mathbf{q}^j
-\frac{1}{3}\left|\mathbf{q}\right|^2\delta^{ij}\right) \\ & \qquad\qquad + v_{5,n} \left( \mathbf{k}^i\mathbf{k}^j
-\frac{1}{3}\left|\mathbf{k}\right|^2\delta^{ij}\right)\Biggr] \\ &  \qquad\qquad\qquad \times
\left(2 \frac{\hat{\mathcal{G}}_1^{ia}\hat{\mathcal{G}}_2^{ja}}{N_c^2}\right)
\left( \frac{\hat{\Lambda}_1 \cdot \hat{\Lambda}_2}{N_c^2}\right)^n,
\end{split}
\end{align}
where $\hat{\Lambda}_1 \cdot \hat{\Lambda}_2 = \hat{\Lambda}_{\gamma\alpha}^M \hat{\Lambda}_{\delta\beta}^M$ and correspondigly for $\hat{\mathcal{S}}$, $\hat{\mathcal{T}}$, and $\hat{\mathcal{G}}$. The range of $\alpha,\dots,\delta$ depends on which internal quantum number they describe and on the representation the involved states belong to. In this potential, the coefficients $v_{k,n},k=0,\dots,5$ are scalar functions of $|\mathbf{q}|^2$ and
$|\mathbf{k}|^2$ and related to the $h_{stu}$ of Eq.~\eqref{eq:HartreeHamil} up to some
unimportant normalization factors and after separating out explicit factors of $\mathbf{q}$
and $\mathbf{k}$ guaranteeing the right behavior under parity, time reversal, and rotational
symmetry. These functions in general are of $\order{1}$ in the large-$N_c$ power counting,
but in the case of terms proportional to $(\mathbf{q} \times \mathbf{k})^i$ a $1/N_c$ suppression
is expected due to Eq.~\eqref{eq:qksuppression}. In Eq.~\eqref{eq:HamilExpansion}, the terms of
the first line yield the central part of the two-baryon potential, terms $\sim (\mathbf{q}
\times \mathbf{k})^i$ the spin-orbit interaction, and the terms of the last line the tensor potentials. 

Explicitly performing the expansion up to order $1/N_c$, the Hamiltonian~\eqref{eq:HamilExpansion}
can be further simplified when restricted to the pure octet baryon sector, resulting in
\begin{align}\label{eq:largeNCPot}
\begin{split}
V&_{\B^\alpha\B^\beta\to \B^\gamma\B^\delta} =  N_c  \Biggg\{ \ v_{0,0}\\ &  + v_{0,1}^{(T)}
\frac{\sdot{\hat{\mathcal{T}}_1}{\hat{\mathcal{T}}_2}}{2 N_c^2} + v_{0,1}^{(S)}
\frac{\sdot{\hat{\mathcal{S}}_1}{\hat{\mathcal{S}}_2}}{3 N_c^2} + 2 v_{0,1}^{(G)} \frac{\sdot{\hat{\mathcal{G}}_1}{\hat{\mathcal{G}}_2}}{N_c^2} \\
& + \Biggg[ \left(v_{1,0} + v_{1,1}^{(T)}
\frac{\sdot{\hat{\mathcal{T}}_1}{\hat{\mathcal{T}}_2}}{2N_c^2} \right) \frac{\left(\hat{\mathcal{S}}^i_1 + \hat{\mathcal{S}}_2^i\right)}{\sqrt{3}N_c} \\ & \qquad + v_{2,0} \frac{\left(\hat{\mathcal{G}}_2^{ia}\hat{\mathcal{T}}^a_1 + \hat{\mathcal{G}}_1^{ia}\hat{\mathcal{T}}_2^a
\right)}{N_c^2} \Biggg] \left(\mathbf{q}\times\mathbf{k}\right)^i \\
& + \Biggl[ v_{4,0} \left( \mathbf{q}^i\mathbf{q}^j-\frac{1}{3}\left|\mathbf{q}\right|^2\delta^{ij}\right)
  + v_{5,0} \left( \mathbf{k}^i\mathbf{k}^j-\frac{1}{3}\left|\mathbf{k}\right|^2\delta^{ij}\right)\Biggr] \\ &\qquad\times
\frac{2\left(\hat{\mathcal{G}}_1^{ia}\hat{\mathcal{G}}_2^{ja}\right)}{N_c^2}\Biggg\} + \order{1/N_c^3} .
\end{split}
\end{align}
At this point, this may be compared to a rather generic, but merely symbolic formulation of the SU(3) baryon-baryon potential with flavor labels $a\dots d$, which can be written as
\begin{align}
\begin{split}\label{eq:symbolicPot}
V_{B^aB^b\to B^cB^d} =&\ \phantom{+} V_0^0 + V_\sigma^0 \sdot{\boldsymbol{\sigma}_1}{\boldsymbol{\sigma}_2}\\ & + V_\text{LS}^0  \sdot{\mathbf{L}}{\mathbf{S}} + V_\text{T}^0 S_{12} \\
 & + V_0^1 \rho_0^{abcd} + V_\sigma^1 \sdot{\boldsymbol{\sigma}_1}{\boldsymbol{\sigma}_2} \rho_\sigma^{abcd} \\ & + V_\text{LS}^1  \sdot{\mathbf{L}}{\mathbf{S}} \rho_\text{LS}^{abcd} + V_\text{T}^1  S_{12} \rho_\text{T}^{abcd} , \\
\end{split}
\end{align}
where
\begin{equation}\label{eq:S12}
S_{12}(\mathbf{\hat{r}}) = 3 \sdot{\mathbf{\hat{r}}}{\boldsymbol{\sigma}_1}\sdot{\mathbf{\hat{r}}}{\boldsymbol{\sigma}_2} - \sdot{\boldsymbol{\sigma}_1}{\boldsymbol{\sigma}_2}
\end{equation}
with $\mathbf{\hat{r}} = \mathbf{r}/|\mathbf{r}|$, and the $\rho^{abcd}_{\left\{0,\sigma,\text{LS},\text{T}\right\}}$ represent some appropriate structure in accordance with SU(3) flavor symmetry not important at this stage.\footnote{Note that in actual potentials derived in the context of baryon chiral perturbation theory the potentials $V^0_{\left\{0,\sigma,\text{LS},\text{T}\right\}}$ do usually not appear isolated but are incorporated into some structures similar to the $\rho^{abcd}_{\left\{0,\sigma,\text{LS},\text{T}\right\}}$, see, e.\,g., the potential Eq.~\eqref{eq:VLOcont}.} Here, we have deliberately mimicked the generic nucleon-nucleon potential given in Ref.~\cite{Kaplan:1996rk} in order to faciliate the comparision. For the nucleon-nucleon interaction, the $\rho^{abcd}_{\left\{0,\sigma,\text{LS},\text{T}\right\}}$ are simply given by $\sdot{\boldsymbol{\tau}_1}{\boldsymbol{\tau}_2}$, with $\boldsymbol{\tau}$ being the isospin operator. What the authors of Ref.~\cite{Kaplan:1996rk} have shown is that in this case only $V_0^0$, $V_\sigma^1$, and $V_\text{T}^1$ are of leading $\order{N_c}$, while all other contributions are of $\order{1/N_c}$. Comparing Eq.~\eqref{eq:symbolicPot} with Eq.~\eqref{eq:largeNCPot} taking account of the scalings given in Eq.~\eqref{eq:TGscalings}, one finds for the SU(3) baryon-baryon interaction considering baryons of strangeness of $\order{1}$
\begin{align}
\begin{split}\label{eq:Vscalings}
V_0^0 & \sim V_0^1 \sim V_\sigma^1 \sim V_\text{T}^1\sim N_c , \\
V_\sigma^0 & \sim V_\text{LS}^0 \sim V_\text{LS}^1 \sim V_\text{T}^0 \sim 1/N_c ,
\end{split}
\end{align}
which is basically the same as for the nucleon-nucleon case except for the lifting of $V_0^1$, which deserves explanation. It has been noted several times
\cite{Dashen:1993as,Jenkins:1995gc,Luty:1993fu,Dashen:1993jt,Dashen:1994qi,Kaplan:1996rk}
that the large-$N_c$ analysis of baryons is more intricate in comparison to the large-$N_c$
analysis of nucleons due to the more complicated scalings of Eq.~\eqref{eq:TGscalings}. This
mainly affects terms $\sim\sdot{\hat{\mathcal{T}}_1}{\hat{\mathcal{T}}_2}$ which in the
corresponding two-nucleon potential are suppressed by a relative factor of $1/N_c^2$ but in
general are not suppressed in the baryon-baryon case, leading to the lifting of $V_0^1$.  On the other hand, considering the ``hidden'' $1/N_c$ suppression due to Eq.~\eqref{eq:qksuppression},
the spin-orbit potentials $V_\text{LS}^i$ are still suppressed by a relative $\order{1/N_c^2}$ as in this case the more complex scaling of $\hat{\mathcal{G}}_{2/1}^{ia}
\hat{\mathcal{T}}^a_{1/2} \sim N_c$ is unambiguous due to the summation over the flavor index.

{Note that in the most general case the baryon-baryon potential Eq.~\eqref{eq:symbolicPot} can also have an antisymmetric spin-orbit term $\sim \mathbf{L}\cdot\left(\boldsymbol{\sigma}_1-\boldsymbol{\sigma}_2\right)$ \cite{Haidenbauer:2013oca}. This force describing spin singlet-triplet transitions is absent in isospin-symmetric nucleon-nucleon potentials but is in accordance with SU(3) symmetry. However, in the large-$N_c$ case this contribution comes with the same suppressions that also showed up in the $V_\text{LS}^i$ case above due to Eq.~\eqref{eq:qksuppression}. As none of the contributions that we discuss in the following sections does actually generate such antisymmetric spin-orbit interactions, this term is excluded from the analysis and from Eq.~\eqref{eq:symbolicPot}.}

{We further remark that the large-$N_c$ results for the potential are not RG-invariant
and that there is a preferred scale, see e.g. Ref.~\cite{Lee:2020esp} (and references therein).
However, the extraction of this preferred scale as discussed in the nucleon-nucleon case 
\cite{Lee:2020esp} can not be answered here as corresponding data are either absent or too imprecise.}

Before heading to the analysis of the baryon-baryon interaction in chiral perturbation theory, we note that for the matching to the $N_c=3$ case it is convenient to apply the rules established in Ref.~\cite{Lutz:2001yb}, that is
\begin{align}\label{eq:largeNcrules}
\begin{split}
\hat{\mathcal{S}}^i |s, a) & = \frac{1}{2}\sigma_{ss^\prime}^{(i)} |s^\prime, a), \\
\hat{\mathcal{T}}^a |s, b) & = i f^{abc} |s, c), \\
\hat{\mathcal{G}}^{ia} |s, b) & = \sigma_{ss^\prime}^{(i)} t^{abc} |s^\prime, c),
\end{split}
\end{align}
where we have introduced the abbreviation
\begin{equation}
t^{abc} = \frac{1}{2} d^{abc} + \frac{i}{3}f^{abc}
\end{equation}
and $f^{abc}$ and $d^{abc}$ are the two rank three tensors of the flavor SU(3) algebra, see also App.~\ref{app:su3}.

\section{Baryon-baryon interaction in chiral perturbation theory: Contact terms}\label{sec:bbchiralcontact}
\subsection{Chiral power counting}
The leading order (LO) baryon-baryon interaction has been investigated in
\cite{Polinder:2006zh,Polinder:2007mp} which has been extended up to next-to-leading order (NLO)
in Refs.~\cite{Petschauer:2013uua,Haidenbauer:2013oca,Haidenbauer:2015zqb,Haidenbauer:2019boi}. More recently, also
the next-to-next-to-leading order (NNLO) case has been studied \cite{Haidenbauer:2023qhf}.

\begin{figure*}
\centering
\includegraphics[width=0.8\textwidth]{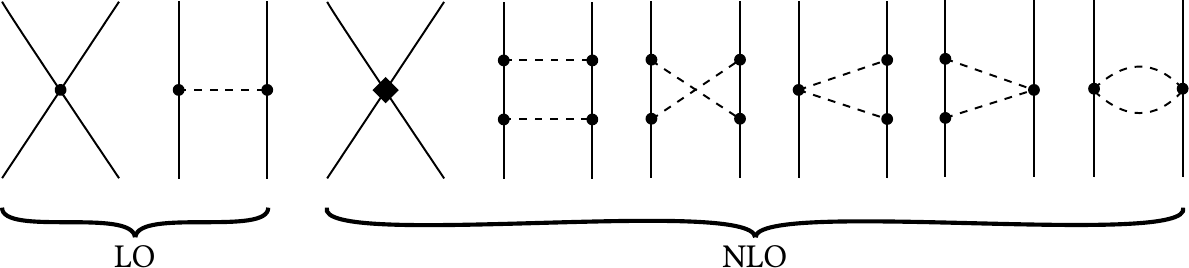}
\caption{Feynman diagrams of the baryon-baryon interaction in chiral perturbation theory
up to next-to-leading order. Solid lines denote octet baryons and dashed lines pseudoscalar
mesons. Dots represent leading order vertices, whereas the diamond denotes a next-to-leading order
contact interaction vertex.}
\label{fig:diagrams}
\end{figure*}

The chiral power counting can be expressed by assigning a chiral order $\order{q^\nu}$,
where $q$ denotes a small momentum or mass. For the baryon-baryon interaction the power
counting  is given by \cite{Weinberg:1990rz,Polinder:2006zh}
\begin{equation}\label{eq:chiralpower}
\nu = 2L + \sum_i v_i\Delta_i,\qquad\Delta_i = d_i +\frac{1}{2} b_i - 2,
\end{equation}
where $v_i$ is the number of vertices of dimension $\Delta_i$ and $L$ is the number of independent
pseudo-Nambu-Goldstone boson loop momenta. The vertex dimension $\Delta_i$ depends on the number
of interacting baryons $b_i$ and the number of pseudo-Nambu-Goldstone boson masses/derivatives $d_i$
at the vertex. In chiral perturbation theory, the pseudo-Nambu-Goldstone bosons are the pseudoscalar
mesons entering the meson-baryon Lagrangian. At LO, corresponding to $\order{q^0}$, only interactions with
$L=0$ (no loops) and $\Delta_i = 0$ contribute corresponding to contributions from leading order contact
interactions ($b_i = 4, d_i=0$) or one-meson exchange (with leading order meson-baryon-baryon vertices,
$b_i=2,d_i=1$; Fig.~\ref{fig:diagrams} left). At NLO, additional contact interactions and two-meson exchange diagrams can contribute (Fig.~\ref{fig:diagrams} right). 

\subsection{Leading Order Contact Interactions}\label{sec:leadingordercontact}
We start with the leading order contribution from contact interactions which is the only
contribution studied in Ref.~\cite{Liu:2017otd}. Let $\Gamma_i$ collectively denote the
elements of the Clifford algebra
\begin{equation}\label{eq:Gammai}
\Gamma_i \in \{\mathbbm{1}, \gamma_\mu, \sigma_{\mu\nu}, \gamma_\mu\gamma_5, \gamma_5\}
\end{equation}
and $B$ the SU(3) baryon octet,
\begin{equation}\label{eq:baryonoctet}
B=\frac{1}{\sqrt{2}}\sum_{a=1}^8 \lambda^a B^a
\end{equation}
then the leading order contact interaction terms corresponding to $\order{q^0}$ in the chiral
power counting read \cite{Polinder:2007mp}
\begin{align}
\begin{split}\label{eq:LOcontLagrangian}
\mathcal{L}_\text{BB}^\text{LO} = & \ \phantom{+} C_i^{(1)} \tr{\bar{B}_\sigma\bar{B}_\tau\left(\Gamma_i B
\right)_\tau \left(\Gamma_i B \right)_\sigma}\\
& + C_i^{(2)} \tr{\bar{B}_\sigma\left(\Gamma_i B \right)_\sigma\bar{B}_\tau\left(\Gamma_i B \right)_\tau }\\
& + C_i^{(3)} \tr{\bar{B}_\sigma\left(\Gamma_i B \right)_\sigma}\tr{\bar{B}_\tau\left(\Gamma_i B
\right)_\tau },
\end{split}
\end{align}
where the $C_i^{(j)}$ are the 15 low-energy constants (LECs) and the subscripts $\sigma$ and $\tau$
are Dirac indices. Throughout this paper, $\tr{\dots}$ denotes the trace in flavor space.
This Lagrangian can be rewritten in a more compact, componentwise notation
\begin{equation}\label{eq:LOcontLagrangianCompact}
\mathcal{L}_\text{BB}^\text{LO} = C_i^{abcd} \Gamma_i^{\sigma_1\sigma_2} \Gamma_i^{\tau_1\tau_2} \bar{B}_{\sigma_1}^c B_{\sigma_2}^a \bar{B}_{\tau_1}^d B_{\tau_2}^b
\end{equation}
where the $C_i^{abcd},i=1,\dots,5$, are linear combinations of the low-energy constants of the
Lagrangian~\eqref{eq:LOcontLagrangian}, 
\begin{equation}\label{eq:Ciabcd}
C_i^{abcd}  = C_i^{(1)}\lambda^{acdb} + C_i^{(2)}\lambda^{cadb} +C_i^{(3)}\delta^{ca}\delta^{db},
\end{equation}
with $\lambda^{abcd}$ as defined in Eq.~\eqref{eq:htlambda}. Let $\left|\mathbf{p}, s, a\right\rangle$ denote a baryon state with momentum $\mathbf{p}$, spin $s$ and SU(3) flavor index $a$, then the potential between two baryon states in the Born approximation of the Lippmann-Schwinger equation is given by the matrix elements
\begin{align}\begin{split}
V&_{B^aB^b\to B^c B^d} = \\& - \left\langle \mathbf{p}^\prime, s_3, c; -\mathbf{p}^\prime, s_4, d\right| \mathcal{L}^\text{int} \left| \mathbf{p}, s_1, a; -\mathbf{p}, s_2, b\right\rangle
\end{split}\end{align}
where $\mathcal{L}^\text{int}$ is the respective interaction Lagrangian. Using the non-relativistic expansion of the Dirac tensor matrix elements given in
App.~\ref{app:bbpotential}, the two-baryon potential derived from the leading order contact interaction Lagrangian Eq.~\eqref{eq:LOcontLagrangianCompact} can be
brought into a form that shows the general pattern of the spin and momentum dependence
\begin{align}
\begin{split}\label{eq:VLOcont}
V&_{B^aB^b\to B^c B^d}^\text{LO,cont.} = \\& \phantom{+} \left(c_S^{abcd} + c_1^{abcd} \left|\mathbf{q}\right|^2+ c_2^{abcd} \left|\mathbf{k}\right|^2 \right) \delta_{s_3s_1}\delta_{s_4s_2} \\
& + \left(c_T^{abcd} + c_3^{abcd} \left|\mathbf{q}\right|^2+ c_4^{abcd} \left|\mathbf{k}\right|^2 \right) \sdot{\boldsymbol{\sigma}_1}{\boldsymbol{\sigma}_2}\\
& +ic_5^{abcd} \mathbf{S}\cdot\left(\mathbf{q}\times\mathbf{k}\right) \\
& + c_6^{abcd} \sdot{\mathbf{q}}{\boldsymbol{\sigma}_1}\sdot{\mathbf{q}}{\boldsymbol{\sigma}_2}+ c_7^{abcd} \sdot{\mathbf{k}}{\boldsymbol{\sigma}_1}\sdot{\mathbf{k}}{\boldsymbol{\sigma}_2}~,
\end{split}
\end{align}
where $\mathbf{q}$ and $\mathbf{k}$ denote the momentum transfer and momentum sum in the baryon's
center-of-mass frame, see Eq.~\eqref{eq:momenta}, the $\boldsymbol{\sigma}_i$'s are the spin
operators of the involved baryons, and 
\begin{equation}
\mathbf{S} = \frac{1}{2} \left(\boldsymbol{\sigma}_1+\boldsymbol{\sigma}_2 \right)
\end{equation} 
is the total spin operator of the two-baryon system. The coefficients $c_k^{abcd}, k
= S,T,1,\dots,7$, are given by
\begin{align}
\begin{split}\label{eq:ciabcd}
c_S^{abcd} & = C_1^{abcd} + C_2^{abcd}, \\ c_T^{abcd} & = C_3^{abcd} - C_4^{abcd},\\
c_1^{abcd} & = \frac{1}{4m_B^2}\left(C_1^{abcd}-C_3^{abcd} \right), \\ c_2^{abcd} & = \frac{1}{2m_B^2} C_2^{abcd},\\
c_3^{abcd} & = -\frac{1}{4m_B^2}\left(C_2^{abcd}+C_4^{abcd} \right), \\ c_4^{abcd} & = \frac{1}{2m_B^2} C_3^{abcd},\\
c_5^{abcd} & = -\frac{1}{8m_B^2} \left(C_1^{abcd}-3C_2^{abcd}-3C_3^{abcd}-C_4^{abcd} \right), \\
c_6^{abcd} & = \frac{1}{4m_B^2}\left(C_2^{abcd} + C_3^{abcd} + C_4^{abcd} - C_5^{abcd}\right),\\
c_7^{abcd} & = -\frac{1}{2m_B^2}\left(C_3^{abcd}+C_4^{abcd}  \right), 
\end{split}
\end{align}
The potential of Eq.~\eqref{eq:VLOcont} shows the usual constituents of the general
two-baryon potential: the first two lines are the central part of the potential, the
third line corresponds to the spin-orbit force, and the last two terms constitute the tensor potential.
\subsection{Large-\titlemath{N_c} analysis}\label{sec:largenccontact}
From the discussion of the large-$N_c$ potential in Section~\ref{sec:largeNCBB} the overall scaling of the involved terms can readily be established. All terms of the first line in Eq.~\eqref{eq:VLOcont} correspond to the terms $\sim v_{0,0}$ and $\sim v_{0,1}^{(T)}$ of the large-$N_c$ potential Eq.~\eqref{eq:largeNCPot} and are allowed to be of $\order{N_c}$. Therein, the parts $\sim |\mathbf{q}|^2$ and $\sim |\mathbf{k}|^2$ are simply part of the expansion of $v_{0,0}$ and $v_{0,1}^{(T)}$ in the momenta, which was only implicit in Eq.~\eqref{eq:VLOcont}. As can be seen from the explicit $1/m_B^2$ factors in Eq.~\eqref{eq:ciabcd}, these terms are suppressed by a relative power of $1/N_c^2$. Consequently, at leading $\order{N_c}$ only $c_S^{abcd}$ contributes. The same argument also holds for the central spin-spin part from the second line of Eq.~\eqref{eq:VLOcont} with respect to the corresponding terms $\sim v_{0,1}^{(S)}$ and $\sim v_{0,1}^{(G)}$ of the large-$N_c$ potential Eq.~\eqref{eq:largeNCPot}. We therefore have to take a closer look at the leading order contributions $\sim c_S^{abcd}$ and $\sim c_T^{abcd}$.

Starting with $c_S^{abcd}$, one needs to match
\begin{align}\begin{split}
c_S^{abcd} & = C_S^{(1)}\lambda^{acdb} + C_S^{(2)}\lambda^{cadb} +C_S^{(3)}\delta^{ca}\delta^{db}\\& \sim N_c \left( v_{0,0} \delta^{ca}\delta^{db} + v_{0,1}^{(T)}
\frac{\sdot{\hat{\mathcal{T}}_1}{\hat{\mathcal{T}}_2}}{N_c^2}\right) ,
\end{split}\end{align}
where as usual $C_S^{(i)} = C_1^{(i)}+C_2^{(i)}$. The most important thing to note is that the term $\sim v_{0,1}^{(T)}$ does only contribute to the leading order potential for $e=8$ in the summation over $(\hat{\mathcal{T}}^e)_{ca} (\hat{\mathcal{T}}^e)_{db}$ due to the rules of Eq.~\eqref{eq:TGscalings}. This heavily restricts the structures of $\lambda^{abcd}$ in $c_S^{abcd}$ that are allowed at $\order{N_c}$. An easy way to see this is by inspection of the actual values of the structure constants, Eq.~\eqref{eq:appstructureconstants} in Appendix~A, which requires that leading order contributions only appear for $a\ne c$ and $b \ne d$ in $(\hat{\mathcal{T}}^8)_{ca} (\hat{\mathcal{T}}^8)_{db}$. With this knowledge and the explicit form of $\lambda^{abcd}$ given in Eq.~\eqref{eq:lambdaabcd} one finds at leading order in $1/N_c$
\begin{align}\begin{split}
\frac{1}{3} (C_S^{(1)} +C_S^{(2)}) + C_S^{(3)} & = N_c v_{0,0} + \order{1/N_c} ,\\
 (C_S^{(1)} +C_S^{(2)}) & = \order{1/N_c} ,\\
(C_S^{(1)} - C_S^{(2)}) & =-2  N_c v_{0,1}^{(T)}+ \order{1/N_c}
\end{split}\end{align}
or
\begin{align}\begin{split}
\frac{C_S^{(1)}}{C_S^{(2)}} & = - 1\left(1 + \order{1/N_c^2}\right),\\
C_S^{(1)} &  \sim C_S^{(2)} \sim C_S^{(3)} \sim N_c ,
\end{split}\end{align}
which is equivalent to the statement above, that $V_0^0$ and $V_0^1$ in Eq.~\eqref{eq:symbolicPot} are of $\order{N_c}$.

Turning to the central spin-spin part the matching requires
\begin{align}\begin{split}
c_T^{abcd} & = C_T^{(1)}\lambda^{acdb} + C_T^{(2)}\lambda^{cadb} +C_T^{(3)}\delta^{ca}\delta^{db} \\ & \sim N_c \left(v_{0,1}^{(S)}\delta^{ca}\delta^{db}
\frac{\sdot{\hat{\mathcal{S}}_1}{\hat{\mathcal{S}}_2}}{3 N_c^2} + 2 v_{0,1}^{(G)} \frac{\sdot{\hat{\mathcal{G}}_1}{\hat{\mathcal{G}}_2}}{N_c^2} \right) ,
\end{split}\end{align}
where as usual $C_T^{(i)} = C_3^{(i)}-C_4^{(i)}$. The first term $\sim v_{0,1}^{(S)}$ is of $\order{1/N_c}$ and thus subleading, but the second term including the summation $(\hat{\mathcal{G}}^e)_{ca} (\hat{\mathcal{G}}^e)_{db}$ over the index $e=1\dots 8$ is of $\order{N_c}$ for $e=1,2,3$, see Eq.~\eqref{eq:TGscalings}. In particular, the $\hat{\mathcal{G}}^e$ with $e=1,2,3$ are generators of the SU(4) subalgebra of the contracted SU(6) spin-flavor group, meaning that the leading order central spin-spin part respects SU(4) spin-isospin symmetry, but not SU(6) spin-flavor symmetry. SU(6) breaking is thus associated with a suppression of $\order{\epsilon/N_c}$ with $\epsilon\sim m_s/\Lambda_\chi$ being a measure of SU(3) flavor symmetry breaking \cite{Dai:1995zg,Kaplan:1996rk,Jenkins:1998wy}. The matching, which is best performed using Eq.~\eqref{eq:tacetbde} of App.~\ref{app:su3}, yields
\begin{align}\begin{split}
\frac{1}{3} (C_T^{(1)} +C_T^{(2)}) + C_T^{(3)} & = \frac{v_{0,1}^{(S)}}{3N_c} + \order{1/N_c^3},\\
(C_T^{(1)} + C_T^{(2)}) & = N_c v_{0,1}^{(G)}  + \order{1/N_c} ,  \\
(C_T^{(1)} - C_T^{(2)}) & ={\frac{4}{9}}N_c v_{0,1}^{(G)}  + \order{1/N_c}
\end{split}\end{align}
or
\begin{align}\begin{split}
C_T^{(3)} & = - \frac{1}{3} (C_T^{(1)} +C_T^{(2)}) + \order{1/N_c} ,\\ 
C_T^{(1)} &  \sim  C_T^{(2)} \sim C_T^{(3)} \sim N_c~,
\end{split}\end{align}
which is equivalent to the statement above, that $V_\sigma^0$ and $V_\sigma^1$ in Eq.~\eqref{eq:symbolicPot} are of $\order{1/N_c}$ and $\order{N_c}$, respectively. Moreover, we find the ratios
\begin{align}\begin{split}
\frac{C_T^{(2)}}{C_T^{(1)}} & = \frac{5}{13} \left(1+\order{1/N_c^2}\right),\\\frac{C_T^{(3)}}{C_T^{(1)}} & = -\frac{6}{13} \left(1+\order{1/N_c^2}\right).
\end{split}\end{align}
It is thus clear that at leading order in $1/N_c$ only terms $\sim C_{S/T}^{(i)}$ contribute to the contact interaction potential and that each coefficient $C_{S/T}^{(i)}$ individually is of $\order{N_c}$. However, certain linear combinations of these coefficients are suppressed, which reduces the number of free parameters in the leading order large-$N_c$ baryon-baryon potential from six to three.

This result implies that the coefficients of the original Lagrangian, Eq.~\eqref{eq:LOcontLagrangian}, $C_i^{(j)}$ each are of $\order{N_c}$ for $i=1\dots 4$ meaning that any other term $\sim c_k^{abcd}$, $k=1\dots 7$, in the potential Eq.~\eqref{eq:VLOcont} is suppressed by $1/N_c^2$ simply due to the factors $1/m_B^2$, see Eq.~\eqref{eq:ciabcd}. The contact interaction hence reproduces the large-$N_c$ predictions Eq.~\eqref{eq:Vscalings} quite well except for the scaling of $V_{T}^1$, which corresponds to $c_6^{abcd}$ in the contact potential Eq.~\eqref{eq:VLOcont}. As will be shown in Section \ref{sec:bbchiralOMETME}, this seemingly ``missing'' $\order{N_c}$ contribution is added by one-meson exchange diagrams.
\subsection{Consistency check: Hyperon-Nucleon potentials in chiral perturbation theory}\label{sec:hyperonnucleoncheck}
According to the results of the previous section, there are three coefficients of the leading order contact potential that can be eliminated. In particular, we found that up to corrections of $\order{1/N_c}$ we are allowed to replace 
\begin{align}
\begin{split}
C_S^{(2)} &\approx - C_S^{(1)}, \\ C_T^{(2)} &\approx  \frac{5}{13} C_T^{(1)}, \\ C_T^{(3)} &\approx -\frac{6}{13} C_T^{(1)}.
\end{split}
\end{align}
Introducing
\begin{equation}
C_+^{(i)} = C_S^{(i)} + C_T^{(i)}, \qquad C_-^{(i)} = C_S^{(i)} -3 C_T^{(i)}~,
\end{equation}
the hyperon-nucleon potentials are given by \cite{Polinder:2006zh,Polinder:2007mp}
\begin{align}\begin{split}
V_{1S0}^{N\Lambda \to N\Lambda} & = 4\pi \left( \frac{1}{6} C_-^{(1)} + \frac{5}{3} C_-^{(2)} +2 C_-^{(3)}\right) \equiv C_{1S0}^{\Lambda\Lambda},\\
V_{3S1}^{N\Lambda \to N\Lambda} & = 4\pi \left( \frac{3}{2} C_+^{(1)} +  C_+^{(2)} +2 C_+^{(3)}\right) \equiv C_{3S1}^{\Lambda\Lambda},\\
V_{1S0}^{N\Sigma \to N\Sigma} & = 4\pi \left(  2 C_-^{(2)} +2 C_-^{(3)}\right) \equiv C_{1S0}^{\Sigma\Sigma},\\
V_{3S1}^{N\Sigma \to N\Sigma} & = 4\pi \left( -2 C_+^{(2)} +2 C_+^{(3)}\right)\equiv C_{3S1}^{\Sigma\Sigma} ,\\
V_{3S1}^{N\Lambda \to N\Sigma} & = 4\pi \left( -\frac{3}{2} C_+^{(1)} +  C_+^{(2)} \right)\equiv C_{3S1}^{\Lambda\Sigma}.
\end{split}\end{align}
Using the large-$N_c$ predictions given above, one finds
\begin{align}
\begin{split}
\label{eq:YNlargeNC}
C_{1S0}^{\Sigma\Sigma} & \approx  \frac{1}{9}\left(20 C_{1S0}^{\Lambda\Lambda} - 11 C_{3S1}^{\Lambda\Lambda}-7 C_{3S1}^{\Lambda\Sigma}  \right)~, \\
C_{3S1}^{\Sigma\Sigma} & \approx -12 C_{1S0}^{\Lambda\Lambda} +13 C_{3S1}^{\Lambda\Lambda}+9 C_{3S1}^{\Lambda\Sigma}~.
\end{split}
\end{align}
These large-$N_c$ sum rules of the leading order contact terms are indeed fulfilled to a good accuracy as can be seen from Table~\ref{tab:YN}. Especially for small cutoff masses, the agreement is formidable with deviations just within what is expected from $1/N_c$ corrections.  {We note that these sum rules differ from the ones given in~\cite{Liu:2017otd}, from the details given in that paper we were not able to arrive at their results.}
\begin{table*}
\centering
\begin{tabular}{|l||c|c|c||c|c||c|c|}
\hline
Cutoff & $C_{1S0}^{\Lambda\Lambda}$ & $C_{3S1}^{\Lambda\Lambda}$ & $C_{3S1}^{\Lambda\Sigma}$ & \multicolumn{2}{c||}{$C_{1S0}^{\Sigma\Sigma}$} &  \multicolumn{2}{c|}{$C_{3S1}^{\Sigma\Sigma}$} \\ \hline
$550$\,MeV & $-0.0466$ & $-0.0222$ & $-0.0016$ & $-0.0766$ & $\mathbf{-0.0751}$ & $0.2336$ & $\mathbf{0.2562}$ \\
$600$\,MeV & $-0.0403$ & $-0.0163$ & $-0.0019$ & $-0.0763$ & $\mathbf{-0.0682}$ & $0.2391$ & $\mathbf{0.2546}$ \\
$650$\,MeV & $-0.0322$ & $-0.0097$ & $\phantom{-}0.0000$ & $-0.0757$ & $\mathbf{-0.0597}$ & $0.2392$ & $\mathbf{0.2603}$ \\
$700$\,MeV & $-0.0304$ & $-0.0022$ & $\phantom{-}0.0035$ & $-0.0744$ & $\mathbf{-0.0676}$ & $0.2501$ & $\mathbf{0.3677}$ \\
\hline
\end{tabular}
\caption{Comparing best fit hyperon-nucleon potentials from Ref.~\cite{Polinder:2006zh} and corresponding large-$N_c$ predictions (in units of $10^4$ GeV$^{-2}$). The bold values of $C_{1S0}^{\Sigma\Sigma}$ and $C_{3S1}^{\Sigma\Sigma}$ are obtained using the large-$N_c$ sum rules Eq.~\eqref{eq:YNlargeNC}.}
\label{tab:YN}
\end{table*}

\subsection{Next-to leading order contact interactions}\label{sec:bbhigherorders}
Ref.~\cite{Petschauer:2013uua} summerizes all contact contributions up-to-and-including
$\order{q^2}$ in the relativistic approach. Let
\begin{equation}\label{eq:mesonoctet}
\Phi = \sum_{a=1}^8 \lambda^a \Phi^a
\end{equation}
denote the SU(3) pseudoscalar-meson octet such that the building blocks that enter the Lagrangian at this order are given by
\begin{align}
\begin{split}\label{eq:blocks}
u & = \exp\left(i\frac{\Phi}{2F_0}\right),\\
\mathcal{D}_\mu B & = \partial_\mu B +\com{\Gamma_\mu}{B},\\
\Gamma_\mu & = \frac{1}{2} \left(u^\dagger\partial_\mu u + u \partial_\mu u^\dagger \right)
= \frac{1}{8F_0^2}\com{\Phi}{\partial_\mu\Phi} + \order{\Phi^4},\\
u_\mu & = i \left(u^\dagger\partial_\mu u- u \partial_\mu u^\dagger \right)
= -\frac{1}{F_0}\partial_\mu\Phi + \order{\Phi^3} \\
\chi_\pm & = u^\dagger\chi u^\dagger \pm u\chi^\dagger u ,
\end{split}
\end{align}
where $\chi = 2 B_0 \mathcal{M}_q$ is proportional to the diagonal quark mass
matrix $\mathcal{M}_q$ and the parameter $B_0$ is related to the quark condensate. Contributions of $\order{q^1}$ in the chiral
power counting have either the chiral covariant derivative $\mathcal{D}_\mu$ or the chiral
building block $u_\mu$. However, in a non-relativistic expansion,
contributions with $\mathcal{D}_\mu$ are actually relegated to $\order{q^2}$ and contributions with $u_\mu$ add at least one pseudoscalar to the vertex meaning
that diagrams with such vertices must contain at least one loop and hence are
of subleading order according to the power counting of Eq.~\eqref{eq:chiralpower}. At $\order{q^2}$, also SU(3) symmetry breaking terms stemming from explicit insertions of the quark mass matrix do appear. Here, only terms with
direct insertions of $\chi$ are relevant, as terms with $\chi_{-}$ are of $\order{q^3}$ in the non-relativistic limit, and any appearances of pseudoscalars in $\chi_+$ are dropped anyway for pure contact interactions. The corresponding Lagrangian is hence given by~\cite{Petschauer:2013uua}
\begin{align}
\begin{split}
\mathcal{L}_\text{BB}^\text{NLO} = & \ \phantom{+} \tilde{C}_i^{(1)} \tr{\bar{B}_\sigma\chi\left(\Gamma_i B \right)_\sigma \bar{B}_\tau\left(\Gamma_i B \right)_\tau} \\
& + \tilde{C}_i^{(2)} \tr{\bar{B}_\sigma\left(\Gamma_i B \right)_\sigma\chi \bar{B}_\tau\left(\Gamma_i B \right)_\tau} \\
& + \tilde{C}_i^{(3)}\Bigl( \tr{\bar{B}_\sigma\chi\bar{B}_\tau\left(\Gamma_i B \right)_\tau \left(\Gamma_i B \right)_\sigma} \\ & \qquad\qquad + \tr{\bar{B}_\sigma\bar{B}_\tau\left(\Gamma_i B \right)_\tau\chi \left(\Gamma_i B \right)_\sigma} \Bigr) \\
& + \tilde{C}_i^{(4)} \tr{\bar{B}_\sigma\bar{B}_\tau\chi\left(\Gamma_i B \right)_\tau \left(\Gamma_i B \right)_\sigma} \\
& + \tilde{C}_i^{(5)} \tr{\bar{B}_\sigma\bar{B}_\tau\left(\Gamma_i B \right)_\tau \left(\Gamma_i B \right)_\sigma\chi} \\
& + \tilde{C}_i^{(6)} \tr{\bar{B}_\sigma\left(\Gamma_i B \right)_\sigma\chi}\tr{\bar{B}_\tau\left(\Gamma_i B \right)_\tau} \\
& + \tilde{C}_i^{(7)} \Bigl(\tr{\bar{B}_\sigma\chi}\tr{\left(\Gamma_i B \right)_\sigma \bar{B}_\tau\left(\Gamma_i B \right)_\tau}\\ & \qquad\qquad + \tr{\bar{B}_\sigma\left(\Gamma_i B \right)_\sigma \bar{B}_\tau}\tr{\left(\Gamma_i B \right)_\tau\chi}\Bigr) , \\
\end{split}
\end{align}
where we use the tilde to distinguish the new LECs from the LO ones. In this context, it
is convenient to decompose $\chi$ into SU(3) symmetric and isospin and SU(3) violating parts
\begin{equation}
\chi = 2B_0\mathcal{M}_q=M^{[0]} \mathbbm{1} + M^{[3]} \lambda_3 + M^{[8]} \lambda_8 
\end{equation}
with
\begin{align}
\begin{split}
M^{[0]} & = \frac{3}{2}\left(M_{\pi^0}^2+M_\eta^2 \right)-\frac{2}{3}\left(M_{\pi^\pm}^2
+ M_{K^\pm}^2 + M_{K^0}^2 \right),\\
M^{[3]} & = M_{K^\pm}^2 - M_{K^0}^2,\\
M^{[8]} & = \frac{1}{\sqrt{3}} \left(2 M_{\pi^\pm}^2 - M_{K^\pm}^2 - M_{K^0}^2  \right),\end{split}
\end{align}
where we have replaced the quark masses and $B_0$ by the leading order SU(3) pseudo-Nambu-Goldstone boson mas\-ses. Introducing 
\begin{align}
\begin{split}\label{eq:Ctildeabcd}
\tilde{C}_i^{abcd} = & \ \phantom{+} M^{[0]} \Biggl\{\left( 2\tilde{C}_i^{(3)}+\tilde{C}_i^{(4)}+\tilde{C}_i^{(5)} \right)\lambda^{cdba}\\ & \qquad +\left(\tilde{C}_i^{(1)}+\tilde{C}_i^{(2)} \right)\lambda^{cadb} +\tilde{C}_i^{(6)}\delta^{ca}\delta^{db} \Biggr\} \\
 & + M^{[3]} \Biggl\{ \tilde{C}_i^{(1)} \lambda^{c3adb} + \tilde{C}_i^{(2)} \lambda^{ca3db} \\ & \qquad +\tilde{C}_i^{(3)} \left[ \lambda^{cdb3a}+\lambda^{c3dba}\right] + \tilde{C}_i^{(4)} \lambda^{cd3ba} \\ & \qquad + \tilde{C}_i^{(5)} \lambda^{cdba3} + \tilde{C}_i^{(6)} \delta^{bd}h^{ca3}\\ & \qquad  + \tilde{C}_i^{(7)} \left[\delta^{c3}h^{adb}+\delta^{b3}h^{cad}\right] \Biggr\} \\ &  + \left([3] \to [8]\right)
\end{split}
\end{align}
with $\lambda^{a_1a_2\dots a_i}$ and $h^{abc}$ as defined in Eq.~\eqref{eq:htlambda}, this next-to-leading order Lagrangian can be rewritten in exactly the same way as the leading order Lagrangian Eq.~\eqref{eq:LOcontLagrangianCompact}
\begin{equation}
\mathcal{L}_\text{BB}^\text{NLO} = \tilde{C}_i^{abcd} \Gamma_i^{\sigma_1\sigma_2} \Gamma_i^{\tau_1\tau_2} \bar{B}_{\sigma_1}^c B_{\sigma_2}^a \bar{B}_{\tau_1}^d B_{\tau_2}^b ,
\end{equation}
with the only difference being that while $C_i^{abcd}$ is symmetric under the exchange of the index pairs $C_i^{abcd} = C_i^{cdab}$, this does not apply to $\tilde{C}_i^{abcd}$. The resulting contributions to the potential are thus of the same form as Eq.~\eqref{eq:VLOcont} with any $c_i^{abcd}$ replaced by their respective counterparts carrying the tilde, and these $\tilde{c}_i^{abcd}$ being set just analogous to Eq.~\eqref{eq:ciabcd}.

The terms $\propto M^{[3]}$ and $M^{[8]}$ violate SU(3) flavor symmetry and there is no matching term in the leading order large-$N_c$ potential Eq.~\eqref{eq:largeNCPot} meaning that any $\tilde{C}_i^{(j)}$ is of subleading order $\order{\epsilon/N_c}$ with $\epsilon$  measuring the SU(3) flavor symmetry breaking~\cite{Kaplan:1996rk}, as noted before.

It is, however, possible to reduce the number of free parameters $\tilde{C}_i^{(j)}$ to leading order in $1/N_c$. This can be seen by assuming SU(3) flavor symmetry, because in this case the terms $\propto M^{[3]}$ and $M^{[8]}$ vanish. The tensors $\tilde{C}_i^{abcd}$ which then are simply $\propto M^{[0]}$ structurally match the $C_i^{abcd}$ of Eq.~\eqref{eq:Ciabcd}. Consequently, the large-$N_c$ rules found in Section~\ref{sec:largenccontact} can readily be translated for the $\tilde{C}_i^{abcd}$ resulting in
\begin{align}\begin{split}
\frac{\tilde{C}_S^{(1)}+\tilde{C}_S^{(2)}}{2\tilde{C}_S^{(3)}+\tilde{C}_S^{(4)}+\tilde{C}_S^{(5)}} & = - 1 \left(1+\order{1/N_c^2}\right),\\
\frac{\tilde{C}_T^{(1)}+\tilde{C}_T^{(2)}}{2\tilde{C}_T^{(3)}+\tilde{C}_T^{(4)}+\tilde{C}_T^{(5)}} & = \frac{5}{13} \left(1+\order{1/N_c^2}\right),\\
\frac{\tilde{C}_T^{(6)}}{ 2\tilde{C}_T^{(3)}+\tilde{C}_T^{(4)}+\tilde{C}_T^{(5)}} & = -\frac{6}{13} \left(1+\order{1/N_c^2}\right).
\end{split}\end{align}
It has been noted in Ref.~\cite{Haidenbauer:2013oca} that currently it is almost impossible to reliably fix these LECs from experimental data. Although the large-$N_c$ analysis leads to an effective reduction of the LECs, this task still seems impracticable. Instead, one might just absorb the higher order contact terms into the leading order LECs which is entirely reliable from a large-$N_c$ viewpoint. The parts $\propto M^{[0]}$ in Eq.~\eqref{eq:Ctildeabcd} obviously constitute just constant shifts to the leading order LECs while the other contributions lead to $\order{\epsilon/N_c}$ corrections.

\section{Baryon-baryon interaction in chiral perturbation theory: Meson-exchange}\label{sec:bbchiralOMETME}
\subsection{One-meson exchange}\label{sec:bbchiralOME}
According to the chiral power counting, Eq.~\eqref{eq:chiralpower}, one-meson exchange (OME) countributions are of the same order as the leading order contact contributions. The leading order meson-baryon Lagrangian reads
\begin{align}
\begin{split}\label{eq:LOPhiBLagr}
\mathcal{L}_{B\Phi}^\text{LO} =& \tr{\bar{B}(i\gamma^\mu\mathcal{D}_\mu-m_0)B}\\&-\frac{D}{2}\tr{\bar{B}\gamma^\mu\gamma_5\acom{u_\mu}{B}}-\frac{F}{2}\tr{\bar{B}\gamma^\mu\gamma_5\com{u_\mu}{B}}
\end{split}
\end{align}
with the building blocks as in Eq.~\eqref{eq:blocks}.
Here, $m_0$ is the baryon octet mass in the three-flavor chiral limit, $D$ and $F$ are coupling
constants related to the axial-vector couplig $g_A = D+F$, and $F_0$ is the pseudoscalar-meson
decay constant in the chiral limit. From the Lagrangian~\eqref{eq:LOPhiBLagr} one can derive
the baryon-baryon-meson (BB$\Phi$) interaction Lagrangian
\begin{equation}
\mathcal{L}_{BB\Phi }^\text{LO} = g_{BB\Phi}^{abc} \bar{B}_b \gamma^\mu\gamma_5 \partial_\mu\Phi_c B_a
\end{equation}
with the general coupling in the SU(3) Gell-Mann basis
\begin{equation}
g_{BB\Phi}^{abc} = \frac{1}{F_0}\left(D d^{abc} + i F f^{abc} \right) .
\end{equation}
The resulting one-meson exchange potential is then given by
\begin{align}
\begin{split}\label{eq:1PhiEPot}
V&_{B^aB^b\to B^c B^d}^\text{OME} = \\&-g_{BB\Phi}^{ace}g_{BB\Phi}^{bde} \frac{1}{|\mathbf{q}|^2
+ M_{\Phi_e}^2-q_0^2}\Bigl\{  \sdot{\mathbf{q}}{\boldsymbol{\sigma}_1}\sdot{\mathbf{q}}
{\boldsymbol{\sigma}_2}\\
& + \frac{q_0}{2m_B} \left[ \sdot{\mathbf{q}}{\boldsymbol{\sigma}_1}\sdot{\mathbf{k}}
{\boldsymbol{\sigma}_2} - \sdot{\mathbf{k}}{\boldsymbol{\sigma}_1}\sdot{\mathbf{q}}
{\boldsymbol{\sigma}_2} \right] \\ 
& + \frac{\sdot{\mathbf{q}}{\mathbf{k}}}{8m_B^2} \left[ \sdot{\mathbf{q}}{\boldsymbol{\sigma}_1}
\sdot{\mathbf{k}}{\boldsymbol{\sigma}_2} + \sdot{\mathbf{k}}{\boldsymbol{\sigma}_1}
\sdot{\mathbf{q}}{\boldsymbol{\sigma}_2} \right]  \Bigr\},
\end{split}
\end{align}
where $M_{\Phi_e}^2$ is the respective meson mass of $\order{N_c^0}$
 \cite{Witten:1979kh} and $q_0$ denotes the energy transfer. A summation over all intermediate mesons $\Phi_e$ is implied. For definiteness we have substituted $m_0$ with $m_B$ as in the large-$N_c$ limit the baryon masses are degenerate up to corrections of relative order $1/N_c^2$. As $q_0 \approx \Delta m_B
+ \sdot{\mathbf{k}}{\mathbf{q}}/(2m_B)$, the first correction term in the second line is of 
$\order{1/N_c^2}$ in relation to the term of the first line, as is the term in the last line, see Eq.~\eqref{eq:momentascalings}, so these terms are suppressed both in terms of a low-energy expansion and in terms of
large-$N_c$ power counting. However, even in the $N_c=3$ case the baryon mass splitting does not
affect interactions with on-shell, equal-mass initial and final baryons, such as $N\Lambda\to N\Lambda$.

It is common practice to split the potential into a central spin-spin part and a tensorial part using $S_{12}$, see Eq.~\eqref{eq:S12}. Neglecting the subleading terms of the potential \eqref{eq:1PhiEPot} and performing this separation of the central and tensorial part, one gets
\begin{align}
\begin{split}
V_{B^aB^b\to B^c B^d}^\text{OME} = & \Biggl\{ - g_{BB\Phi}^{ace}g_{BB\Phi}^{bde} \frac{1}{3} \frac{|\mathbf{q}|^2}{|\mathbf{q}|^2
+ M_{\Phi_e}^2} \sdot{\boldsymbol{\sigma}_1}{\boldsymbol{\sigma}_2} \\
& \quad - g_{BB\Phi}^{ace}g_{BB\Phi}^{bde} \frac{1}{|\mathbf{q}|^2
+ M_{\Phi_e}^2} \\ &\times\left[ \sdot{\mathbf{q}}{\boldsymbol{\sigma}_1}\sdot{\mathbf{q}}{\boldsymbol{\sigma}_2} -\frac{1}{3} |\mathbf{q}|^2\sdot{\boldsymbol{\sigma}_1}{\boldsymbol{\sigma}_2} \right] \Biggr\} \\
& \times \left( 1 + \order{\frac{1}{N_c^2}}\right)
\end{split}
\end{align}
where we kept the tensorial part of the second line explicit instead of substituting $S_{12}$. In this form, the potential can directly be compared with the large-$N_c$ potential of Eq.~\eqref{eq:largeNCPot} and it can be seen immediately that the term of the first line corresponds to the large-$N_c$ term $\sim v_{0,1}^{(G)}$ and the terms of the second and third line to the terms $\sim v_{4,0}$. By inspection of the rules Eq.~\eqref{eq:largeNcrules}, it is clear that these terms $\sim \sdot{\hat{\mathcal{G}}_1}{\hat{\mathcal{G}}_2} \sim t^{ace} t^{bde}$, so the large-$N_c$ series requires that $g_{BB\Phi}^{abc} \sim t^{abc}$ which is only possible if $F/D = 2/3 \left( 1 + \order{1/N_c^{2}}\right)$, which of course is a well-known result that has been derived several times before using various approaches, see e.\,g. \cite{Dashen:1993jt}. From $g_A=D+F$ one can hence derive that $D=\tfrac{3}{5}g_A(1+\order{1/N_c^2})$ and $F=\tfrac{2}{5}g_A(1-\order{1/N_c^2})$. Taking $g_A=1.26$, this can be estimated to be $D \approx 0.84$ and $F\approx 0.45$ for the $N_c=3$ case, which is remarkebly close to the values $D=0.81(4)$ and $F=0.44(3)$ that can be derived from the current FLAG Review values for the flavor diagonal axial charges \cite{FlavourLatticeAveragingGroupFLAG:2021npn} -- within errors and corrections of higher order in $1/N_c$.

A viable large-$N_c$ OME potential is thus given by
\begin{align}\begin{split}\label{eq:largeNcOME}
V&_{B^aB^b\to B^c B^d}^{\text{OME, large-}N_c} = \\& - t^{ace}t^{bde} \frac{1}{3} \left(\frac{6 g_A}{5 F_0} \right)^2 \frac{|\mathbf{q}|^2}{|\mathbf{q}|^2
+ M_{\Phi_e}^2} \left[ \sdot{\boldsymbol{\sigma}_1}{\boldsymbol{\sigma}_2} + S_{12}(\mathbf{\hat{q}}) \right]
\end{split}\end{align}
or, equivalently
\begin{align}\begin{split}
V&_{B^aB^b\to B^c B^d}^{\text{OME, large-}N_c} = \\& - t^{ace}t^{bde} \left(\frac{6 g_A}{5 F_0} \right)^2 \frac{1}{|\mathbf{q}|^2
+ M_{\Phi_e}^2}\sdot{\mathbf{q}}{\boldsymbol{\sigma}_1}\sdot{\mathbf{q}}
{\boldsymbol{\sigma}_2} .
\end{split}\end{align}
Finally, comparing this again with the large-$N_c$ potential from the Hartree Hamiltonian Eq.~\eqref{eq:largeNCPot}, this potential must at most scale as $\order{N_c}$. This is indeed the case, as $g_A = \order{N_c}$ (which hence also aplies to $F$ and $D$) and $F_0=\order{\sqrt{N_c}}$. However, the spin-flavor structure reveals that this scaling is only the maximum expectable scaling for terms $\sim \hat{\mathcal{G}}_1^{ie}\hat{\mathcal{G}}_2^{je}$, see Eq.~\eqref{eq:TGscalings}. In particular, this means that there is a hierarchy among the exchange particles, as for $e=1,2,3$ (pions) the potential is of $\order{N_c}$, for $e=4,5,6,7$ (kaons) it is of $\order{1}$, while for $e=8$ ($\eta$) it is suppressed by a factor $1/N_c$. Note that this large-$N_c$ result only applies to baryons with strangeness of $\order{1}$. A similar hierarchy is evident also in terms of the exchange meson masses, as the heavier particles lead to potentials of shorter range. Overall, this justifies the exclusion of the $\eta$ particle from studies of the hyperon-nucleon potential, as has been done, e.\,g., in Ref.~\cite{Polinder:2006zh}.

Now consider the limit of very small momentum transfers, $|\mathbf{q}|\to 0$ such that the OME potential Eq.~\eqref{eq:largeNcOME} varies like $|\mathbf{q}|^2/M_{\Phi_e}^2$ and assume that the meson masses are degenerate, i.\,e. the sum over the intermediate mesons $e$ is independet of $M_{\Phi_e}$. In this case, Eq.~\eqref{eq:tacetbde} from the Appendix~A can be applied, such that -- under the assumption that at least one of the incoming baryons is also present in the final state -- the potential reads
\begin{align}
\begin{split}
V_{B^aB^b\to B^c B^d}^{\text{OME, large-}N_c} \approx &  - \left(\frac{1}{5} \lambda^{acbd}+\frac{1}{25} \lambda^{cabd} -\frac{2}{25} \delta^{ac}\delta^{bd}\right) \\&\times\left( \frac{g_A }{ F_0 M_{\Phi}} \right)^2 |\mathbf{q}|^2 \left[ \sdot{\boldsymbol{\sigma}_1}{\boldsymbol{\sigma}_2} + S_{12}(\mathbf{\hat{q}}) \right] ,
\end{split}\end{align}
where the expression in the first parentheses matches the structure of the $C_i^{abcd}$, Eq.~\eqref{eq:Ciabcd} of the leading order contact potential, which was given in Eq.~\eqref{eq:VLOcont}. Evidently, the large-$N_c$ OME potential in this limit can thus be incorporated into the coefficients $c_3^{abcd}$ and $c_6^{abcd}$. However, while formally these terms are allowed to be of $\order{N_c}$, it turned out that -- within the pure contact interaction -- they are actually suppresed by a relative factor of $1/N_c^2$ due to the $1/m_B^2$ factor in the definitions of $c_3^{abcd}$ and $c_6^{abcd}$. Incorporating the large-$N_c$ OME potential, these parts of the potential are finally lifted to the allowed $\order{N_c}$ scaling. This implies that a decent description of the baryon-baryon potential should at least include leading order contact terms and leading order OME contributions.

\subsection{Two-meson exchange}\label{sec:bbchiralTME}
\subsubsection{General remarks}\label{sec:TMEgeneral}
Recent studies of hyperon-nucleon interactions, see e.\,g. Ref.~\cite{Haidenbauer:2013oca,Haidenbauer:2015zqb}, also add two-meson exchange (TME) contributions which also appear at next-to-leading order, see Fig.~\ref{fig:diagrams}. These contributions correspond to box, crossed-box, triangle and football Feynman diagrams, and we denote the corresponding potentials by $V^{\Box}$, $V^{\bowtie}$, $V^{\vartriangleright}$, $V^{\vartriangleleft}$, and $V^{\football}$, respectively. The written-out results are summarized in the Appendix of Ref.~\cite{Haidenbauer:2013oca}. These contributions require dimensional regularization introducing a scale $\lambda$ and the divergent terms are absorbed by contact term LECs of the same chiral order. Here, we will study their large-$N_c$ behavior.

The triangle and football diagrams require the insertion of the leading order BB$\Phi\Phi$ vertex which can be derived from Eq.~\eqref{eq:LOPhiBLagr} and is given by
\begin{equation}
-i g_{BB\Phi\Phi}^{abij} \gamma_\mu \left(q^\mu_1 + q^\mu_2\right)
\end{equation}
with $q_1$ (incoming) and $q_2$ (outgoing) being the four-mo\-menta of the mesons and the coupling tensor is given by
\begin{equation}
g_{BB\Phi\Phi}^{abij} = \frac{1}{2F_0^2} f^{abe}f^{ije},
\end{equation}
where $a,b$ ($i,j$) are flavor indices for the incoming and outgoing baryons (mesons). 

\begin{figure}
\centering
\includegraphics[width=0.25\textwidth]{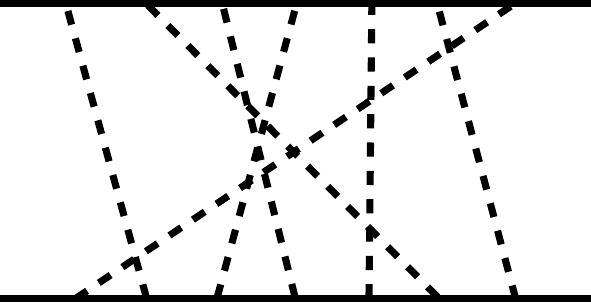}
\caption{Example of a seven meson exchange diagram of the baryon-baryon interaction. Intermediate mesons are non-interacting.}
\label{fig:7ME}
\end{figure}

In general the couplings $g_{BB\Phi^n}$ with an even number $n$ of mesons derived from the first term of the Lagrangian Eq.~\eqref{eq:LOPhiBLagr} are $\propto 1/F_0^n$ and thus of $\order{N_c^{-n/2}}$. On the other hand, the couplings $g_{BB\Phi^n}$ with an odd number $n$ of mesons derived from the $D$ and $F$ terms of the Lagrangian Eq.~\eqref{eq:LOPhiBLagr} are $\propto g_A/F_0^n$ and thus of $\order{N_c^{1-n/2}}$ which is consistent with what is expected from the large-$N_c$ analysis on the quark-gluon level \cite{Jenkins:1998wy}. It is thus tempting to classify any meson exchange diagram with arbitrary many intermediate mesons by simply assigning these large-$N_c$ scalings to the vertices and counting the powers. This will, however, lead to deceptive results. The easiest way to see this is by considering a diagram with an arbitrary number $m$ of non-interacting intermediate mesons all coupled by simple BB$\Phi$ vertices in any order. An example of such a diagram with seven intermediate mesons is shown in Fig.~\ref{fig:7ME}. Assigning a factor of $\sqrt{N_c}$ at each vertex leads to an overall large-$N_c$ scaling of $\left(\sqrt{N_c}\right)^{2m} = N_c^m$ which is in conflict with the prediction that the baryon-baryon potential can at most scale $\sim N_c$ \cite{Witten:1979kh}. 

In fact, the same problem already arises in the case of nucleon-pion scattering, and in general baryon-meson scattering \cite{Flores-Mendieta:1999ctw,Flores-Mendieta:2000ljq}, and it has been shown that consistency with the large-$N_c$ prediction is preserved by considering the contracted SU($2N_f$) spin-flavor algebra discussed in Sec.~\ref{sec:largeNCBB} including the corresponding degenerate baryon tower \cite{Dashen:1993as}. So on a formal level, the assignment that the exemplary $m$ meson exchange diagram scales as $N_c^m$ is correct when considered in isolation. But it is the contracted spin-flavor symmetry that prevents the overall amplitude from blowing up after including also all possible intermediate baryons from the full baryon tower and after adding up any crossed partner diagrams. The symmetry constraints then must ensure the cancellation of the problematic parts.

For the case of the nucleon-nucleon potential, the authors of Ref.~\cite{Banerjee:2001js} have shown explicitly that this works out as expected at the level of two-boson exchange. In accordance with the statements above, this required the inclusion of intermediate $\Delta$ particles which are the only additional members of the spin-isospin tower besides the nucleons (at $N_c=3$). For the present case of $N_f=3$ this means that the integration of decuplet baryons as intermediate states is mandatory. This constrains the value of the octet-decuplet-meson coupling to the known large-$N_c$ value as will be shown in this section.

Finally, the actual overall maximum large-$N_c$ scaling of an arbitrary $n$-meson exchange diagram can be determined by the maximum allowed large-$N_c$ scaling of a general $BB\Phi^n$ vertex, which is given by $\order{N_c^{1-n/2}}$ \cite{Jenkins:1998wy}. So instead of assigning $2n$ simple $g_{BB\Phi}$ vertices of $\order{\sqrt{N_c}}$ to a diagram such as the one given in Figure~\ref{fig:7ME}, one just assigns a factor of $\order{N_c^{1-n/2}}$ to each baryon line meaning that $n$-meson exchange contributions count as $\order{N_c^{2-n}}$. Note, that adding more mesons in such diagrams does not only diminish their weight from a large-$N_c$ perspective, but also in terms of the chiral power counting, Eq.~\eqref{eq:chiralpower}, as each additional meson adds another independent pseudoscalar loop momentum.

\subsubsection{Decuplet Lagrangian}
We use the description of the chiral decuplet-octet interaction as presented in \cite{Jenkins:1992pi,Hemmert:1997ye,Sasaki:2006cx,Haidenbauer:2017sws}. The decuplet fields can be collected into a totally symmetric tensor 
\begin{align}
T_{111} & = \Delta^{++}, & & & & \nonumber\\
T_{112} & = \frac{1}{\sqrt{3}}\Delta^{+}, & T_{113} & = \frac{1}{\sqrt{3}}\Sigma^{*+}, &  T_{133} & =\frac{1}{\sqrt{3}} \Xi^{*0},\nonumber \\ 
T_{122} & = \frac{1}{\sqrt{3}} \Delta^{0},& T_{123} & = \frac{1}{\sqrt{6}}\Sigma^{*0}, & T_{233} & = \frac{1}{\sqrt{3}}\Xi^{*-}, \nonumber\\
T_{222} & = \Delta^{-},& T_{223} & = \frac{1}{\sqrt{3}} \Sigma^{*-}, & T_{333} & = \Omega^{-}, \label{eq:decupletfields}
\end{align}
such that the octet-decuplet-meson interaction Lagrangian can be written as
\begin{equation}\label{eq:decupletlagrangian}
\mathcal{L}_{BT\Phi} = \frac{C}{2 F_0} \sum_{i,j,k,m,n=1}^3 \epsilon_{imn} \left(\bar{T}_{ijk} \sdot{\mathbf{S}^\dagger}{\nabla} \Phi_{jm} B_{kn} + \text{h.c.} \right)
\end{equation}
with the spin transition operators
\begin{align}\begin{split}\label{eq:spintransitionoperators}
S_1 & = \frac{1}{\sqrt{2}}\begin{pmatrix}
-1 & 0 & \frac{1}{\sqrt{3}} & 0 \\
0 & - \frac{1}{\sqrt{3}} & 0 & 1
\end{pmatrix} ,\\
S_2 & =  -\frac{i}{\sqrt{2}}\begin{pmatrix}
1 & 0 & \frac{1}{\sqrt{3}} & 0 \\
0 & \frac{1}{\sqrt{3}} & 0 & 1
\end{pmatrix} ,\\
S_3 & = \begin{pmatrix}
0 & \sqrt{\frac{2}{3}} & 0 & 0 \\
0 & 0 & \sqrt{\frac{2}{3}} & 0 
\end{pmatrix} ,
\end{split}
\end{align}
connecting the two-component octet spinors and the four-component decuplet spinors. These obey
\begin{equation}
S_iS_j^{\dagger} = \frac{2}{3}\delta_{ij} - \frac{i}{3} \epsilon_{ijk} \sigma_k . 
\end{equation}
Being spin-$3/2$ particles, the decuplet fields are given by Rarita-Schwinger fields. However, as the present large-$N_c$ analysis allows for an effective static limit approach to the baryon kinematics, we treat them non-relativistically from the beginning. This is in contrast to the previous sections, where the non-relativistic expansions were performed just in the course of the calculations. Therefore, we can now apply the effective BB$\Phi$ vertex functions
\begin{equation}
g_{BB\Phi}^{abc} \sdot{\boldsymbol{\sigma}}{\mathbf{q}}
\end{equation}
with $\mathbf{q}$ being the three-momentum of an incoming meson and the large-$N_c$ coupling constant given by
\begin{equation}
g_{BB\Phi}^{abc} = \frac{6}{5}\frac{g_A}{F_0} t^{abc} 
\end{equation}
as determined in the last section. Of course, another quite natural choice would be to use heavy baryon chiral perturbation theory for both the octet and the decuplet sector (HBCHPT, see Refs.~\cite{Jenkins:1990jv,Jenkins:1992pi}). Either approach leads to the same conclusion when working to leading order in large-$N_c$, but the present choice seems to be best suited for a concise presentation. In this approach we can safely use 
\begin{equation}\label{eq:baryonpropa}
\frac{i}{p^0 - \frac{|\mathbf{p}|^2}{2m_B} + i\epsilon} \left(1 + \order{\frac{1}{N_c}} \right)
\end{equation}
as the common baryon propagator for both octet and decuplet fields.

As in the previous sections, we strive to separate the spinor fields from their SU(3) content by defining appropriate coupling tensors. There are several ways to achieve this, and here we choose a representation that is similar to the decomposition of octet fields as given, e.\,g. in Eqs.~\eqref{eq:baryonoctet} and~\eqref{eq:mesonoctet}, that is
\begin{equation}
T_{ijk} = \sum_{A=1}^{10} T_A \left(\theta^{A}\right)_{ijk},
\end{equation}
where from now on a Latin capital index represents a decuplet flavor index running from one to ten, and the decuplet fields are identified as
\begin{align}
T_1 & = \Delta^{++},  & & & & \nonumber\\
T_2 & = \Delta^+,  & T_5 & = \Sigma^{*+}, & T_8 & = \Xi^{*0},\nonumber \\
T_3 & = \Delta^0,  & T_6 & = \Sigma^{*0}, & T_9 & = \Xi^{*-}, \\
T_4 & = \Delta^-,   & T_7 & = \Sigma^{*-},& T_{10} & = \Omega^-. \nonumber
\end{align}
The Lagrangian above can then be written
\begin{equation}
\mathcal{L}_{BT\Phi} = \left(g_{BT\Phi}^{Aac}\right)^\dagger \bar{T}^A\sdot{\mathbf{S}^\dagger}{\nabla} \Phi^c B^a + \text{h.c.} 
\end{equation}
It is quite inconvenient to explicitly derive the ten $3\times 3 \times 3$ matrices $\theta^A$ from the tensor $T_{ijk}$, Eq.~\eqref{eq:decupletfields}, instead one might define 
\begin{equation}
\Theta_{ijk} = \begin{cases}
1, & \text{if } i=j=k, \\
\frac{1}{\sqrt{6}}, & \text{if } \{i,j,k\} \in \sigma\left(\{1,2,3\}\right), \\
\frac{1}{\sqrt{3}}, & \text{otherwise}
\end{cases}
\end{equation}
where $\sigma$ denotes the permutation group, and the sets
\begin{align}
P_1 & = \{1,1,1\}, & P_2 & = \{1,1,2\},\nonumber\\ P_3 & = \{1,2,2\}, & P_4 & = \{2,2,2\},\nonumber\\ P_5 & = \{1,1,3\}, &
P_6 & = \{1,2,3\}, \\ P_7 & = \{2,2,3\}, & P_8 & = \{1,3,3\}, \nonumber\\ P_9 & = \{2,3,3\}, & P_{10} & = \{3,3,3\}, \nonumber
\end{align}
which are just the independent indices of $T_{ijk}$, Eq.~\eqref{eq:decupletfields}. Then the coupling tensor can be written
\begin{equation}
g_{BT\Phi}^{Aac} = \frac{1}{\sqrt{2}}\frac{C}{2F_0} \sum_{m,n=1}^3 \sum_{\substack{\{i,j,k\}\\ \in \sigma(P_A)}} \epsilon_{imn} \Theta_{ijk} \left(\lambda^c\right)_{mj}\left(\lambda^a\right)_{nk} .
\end{equation}
\subsubsection{Football diagram}
Beginning with the football diagram, the resulting potential only contributes to the central part of the baryon-baryon potential, $V_0^1$ in Eq.~\eqref{eq:symbolicPot}. It can be written as 
\begin{equation}
V^{\football}_{B^aB^b\to B^c B^d} = g_{BB\Phi\Phi}^{acij} g_{BB\Phi\Phi}^{bdij} V^{\football}_0\left(|\mathbf{q}|^2, M_{\Phi_i}, M_{\Phi_j}, \lambda\right),
\end{equation}
where $V^{\football}_0$ is a function of $|\mathbf{q}|^2$, the involved mesons masses $M_{\Phi_i}$ and $M_{\Phi_j}$, and the scale $\lambda$, see Ref.~\cite{Haidenbauer:2013oca} for details. All of these quantities scale as $\order{1}$ in the large-$N_c$ limit, so it is the coupling that solely determines the large-$N_c$ behavior. As $g_{BB\Phi\Phi} \sim \order{1/N_c}$, this potential is of $\order{1/N_c^2}$ and cleary suppressed in comparison to other contributions.

Assuming degenerate meson masses, the implicit sum over the indices $i,j$ can be performed using Eq.~\eqref{eq:appff} which yields
\begin{equation}
V^{\football}_{B^aB^b\to B^c B^d} = \frac{3}{4F_0^4} f^{ace} f^{bde} V^{\football}_0\left(|\mathbf{q}|^2, M_{\Phi}, \lambda\right).
\end{equation}
Relating to the Hartree potential Eq.~\eqref{eq:largeNCPot}, this contribution is part of the unknown expansion of $v_{0,1}^{(T)}$ in the momenta and hence consistent with the predictions.

Moreover, the more general class of ``football'' like diagrams with the same $BB\Phi^n$ vertex at each baryon line is of $\order{N_c^{2-n}}$ if $n$ is odd but of $\order{N_c^{-n}}$ if $n$ is even. Therefore, the large-$N_c$ scaling of even $n$-meson exchange football diagrams in chiral perturbation theory is less than the allowed $\order{N_c^{2-n}}$.

\subsubsection{Triangle diagrams}
\begin{figure}
\centering
\includegraphics[width=0.45\textwidth]{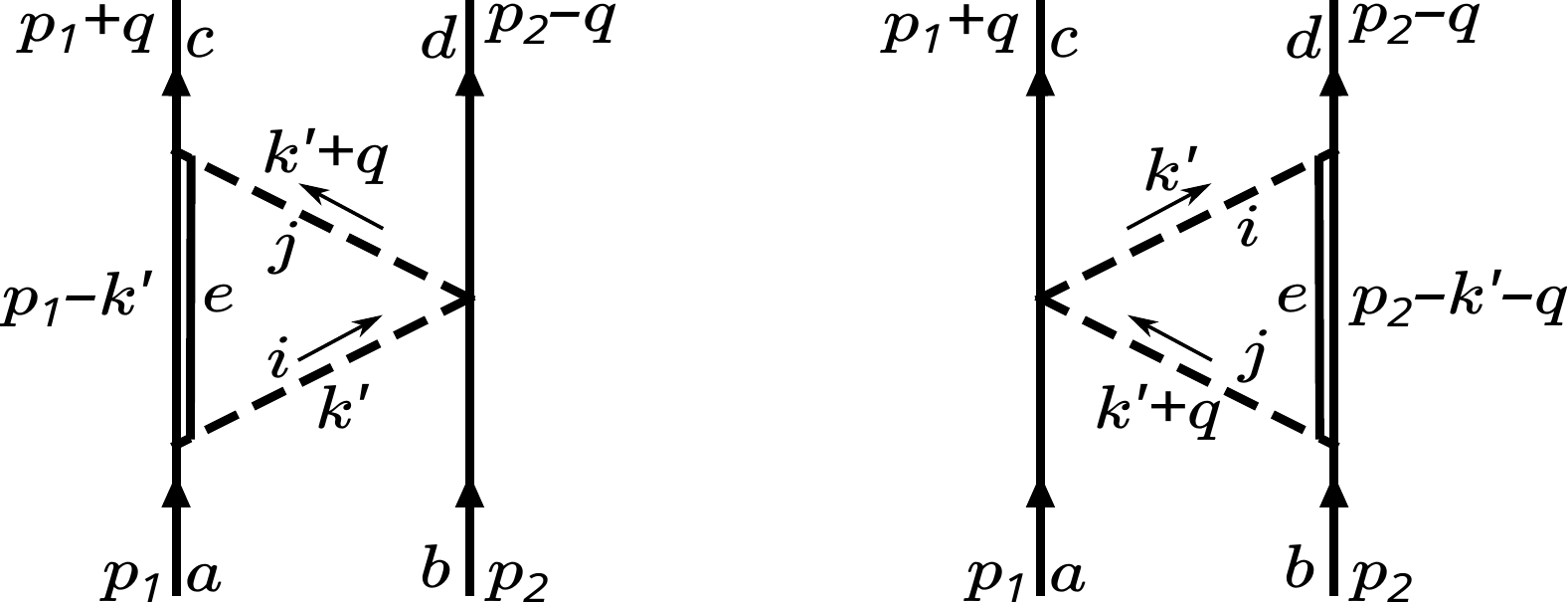}
\caption{Triangle diagrams. Dashed lines denote exchange mesons, solid lines baryons. Double lines denote either octet or decuplet intermediate baryons. {In the latter case, the flavor index $e$ should be replaced by a capital $E$ to indicate a range from 1 to 10.}}
\label{fig:triangles}
\end{figure}
Figure~\ref{fig:triangles} shows collectively the triangle diagrams for both the intermediate octet and decuplet case. As the football diagram, the triangle diagrams contribute to the central potential only and it can be written as a product of three coupling tensors and some function of $|\mathbf{q}|^2$, the meson masses, and the scale $\lambda$
\begin{equation}
V^{\vartriangleright/\vartriangleleft}_{B^aB^b\to B^c B^d} \sim  g_{abcd}^3 V^{\vartriangleright}_0\left(|\mathbf{q}|^2, M_{\Phi_i}, M_{\Phi_j}, \lambda\right) ,
\end{equation} 
where $g_{abcd}^3$ symbollically stands for some appropriate combination of $g_{BB\Phi}$, $g_{BT\Phi}$, and $g_{BB\Phi\Phi}$. The first thing to show is that the function $V^{\vartriangleright}_0$ up to leading order in $1/N_c$ is the same for both triangle diagrams and for both intermediate octet and decuplet. The loop integral involves a non-relativistic baryon propagator as given in Eq.~\eqref{eq:baryonpropa}, two meson propagators, which are the same in any of these diagrams, and some spin-momentum structure from the vertices. Let $\tilde{V}_0^{\vartriangleright/\vartriangleleft}(k^\prime,q,M_{\Phi_i}, M_{\Phi_j})$ collect all of these contributions except for the baryon propagators and the coupling tensors, then the potentials corresponding to the diagrams in Figure~\ref{fig:triangles} are given by
\begin{align}
\begin{split}
V^{\vartriangleright}_0 & = \int \frac{\mathrm d^3k^\prime}{(2\pi)^3}\int\frac{\mathrm dk^\prime_0}{2\pi}\ \frac{\tilde{V}_0^{\vartriangleright}(k^\prime,q,M_{\Phi_i}, M_{\Phi_j})}{-k_0^\prime+\frac{|\mathbf{p}|^2}{2m_B}-\frac{|\mathbf{p}-\mathbf{k}^\prime|^2}{2m_B}}, \\
V^{\vartriangleleft}_0  & = \int \frac{\mathrm d^3k^\prime}{(2\pi)^3}\int\frac{\mathrm dk^\prime_0}{2\pi}\ \frac{\tilde{V}_0^{\vartriangleleft}(k^\prime,q,M_{\Phi_i}, M_{\Phi_j})}{-k_0^\prime-q_0+\frac{|\mathbf{p}|^2}{2m_B}-\frac{|\mathbf{p}+\mathbf{k}^\prime+\mathbf{q}|^2}{2m_B}},
\end{split}
\end{align}
using the center-of-mass momenta $p_1=\left(|\mathbf{p}|^2/(2m_B), \mathbf{p} \right)$ and $p_2=\left(|\mathbf{p}|^2/(2m_B),- \mathbf{p} \right)$. Note that for the sake of brevity we omit factors $(1+\order{1/N_c})$ in this and the following equation below.  Written in this form, it is clear that only the imaginary parts of $\tilde{V}_0^{\vartriangleright/\vartriangleleft}$ contribute to the potential which follows from the Kramers-Kronig relations and the contributions from the baryon propagators are simply given by their {principal} values $\mathcal{P}$
\begin{align}
\begin{split}
V^{\vartriangleright}_0 & = -i\int \frac{\mathrm d^3k^\prime}{(2\pi)^3} \mathcal{P} \left[ \int\frac{\mathrm dk^\prime_0}{2\pi}\ \frac{\Im\left[\tilde{V}_0^{\vartriangleright}(k^\prime,q,M_{\Phi_i}, M_{\Phi_j})\right]}{k_0^\prime}\right], \\
V^{\vartriangleleft}_0  & = -i\int \frac{\mathrm d^3k^\prime}{(2\pi)^3}\mathcal{P} \left[ \int\frac{\mathrm dk^\prime_0}{2\pi}\ \frac{\Im\left[\tilde{V}_0^{\vartriangleleft}(k^\prime,q,M_{\Phi_i}, M_{\Phi_j})\right]}{k_0^\prime}\right] , 
\end{split}
\end{align}
meaning that $V_0^\vartriangleleft = V_0^\vartriangleright$ if $\tilde{V}_0^\vartriangleleft = \tilde{V}_0^\vartriangleright$. To leading order in $1/N_c$ this is indeed the case considering octets and decuplets individually. The only difference is a factor of $2/3$ occuring in the decuplet case stemming from the spin structure in $\tilde{V}_0^{\vartriangleright/\vartriangleleft}$ which can be pulled out and put in front of $V_0^{\vartriangleright/\vartriangleleft}$. The resulting potential is then given by
\begin{align}\begin{split}
V^{\vartriangleright,\vartriangleleft}_{B^aB^b\to B^c B^d} = & -i  \Biggl[  \left(g_{BB\Phi}^{aei}g_{BB\Phi}^{cje}+\frac{2}{3}g_{BT\Phi}^{Eai}g_{BT\Phi}^{Ecj} \right) g_{BB\Phi\Phi}^{bdij} \\ & + \left(g_{BB\Phi}^{bej}g_{BB\Phi}^{die}+\frac{2}{3}g_{BT\Phi}^{Ebj}g_{BT\Phi}^{Edi} \right) g_{BB\Phi\Phi}^{acji} \Biggr] \\ & \qquad\times V^{\vartriangleright}_0\left(|\mathbf{q}|^2, M_{\Phi_i}, M_{\Phi_j}, \lambda\right)\\&\qquad\times \left(1 + \order{\frac{1}{N_c}} \right),
\end{split}\end{align}
where the implicit summation runs from $1\dots 8$ in the case of $i,j,e$, and from $1\dots 10$ in the case of the index $E$. The explicitely spelled out potential $V_0^\vartriangleright$ can be found in the appendix of Ref.~\cite{Haidenbauer:2013oca}. The large-$N_c$ scaling determined from the coupling tensors is given by $\order{1}$, meaning that its contribution to the central potential is more important than the football contribution.

\subsubsection{Box diagrams}
\begin{figure}
\centering
\includegraphics[width=0.45\textwidth]{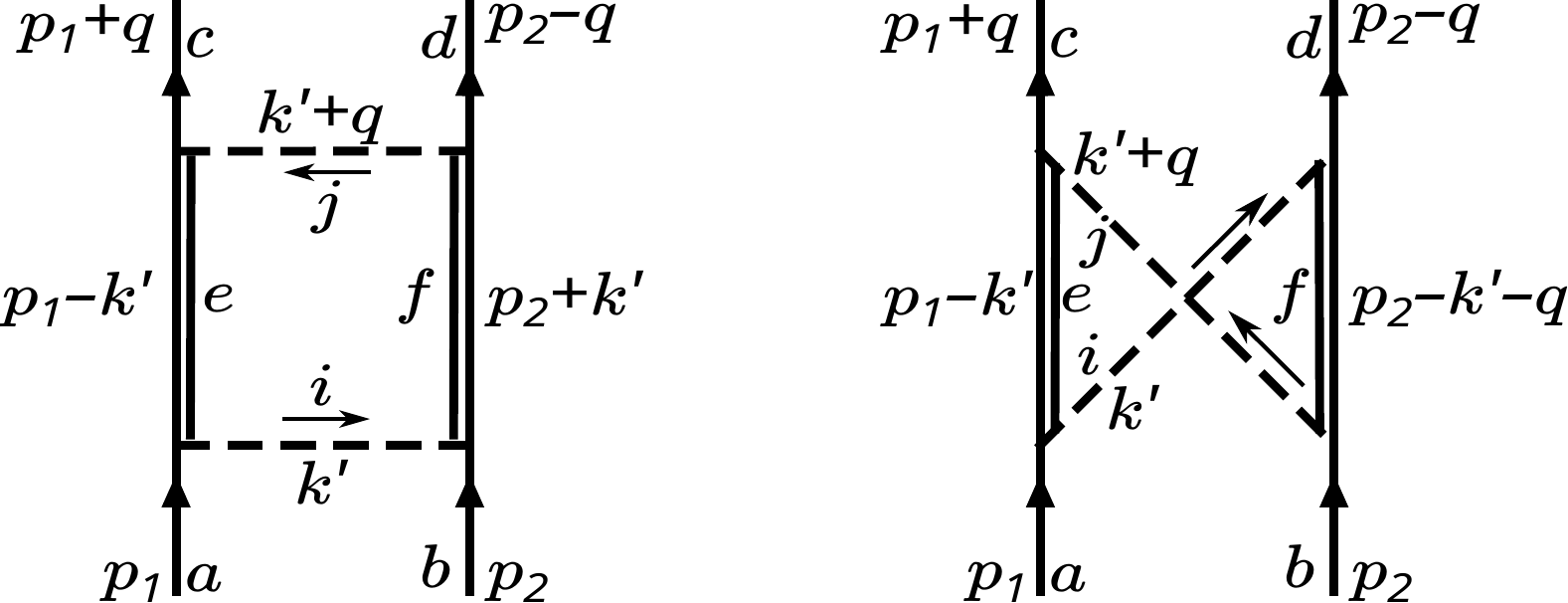}
\caption{Box and crossed box diagrams. Dashed lines denote exchange mesons, solid lines baryons. Double lines denote either octet or decuplet intermediate baryons. {For each intermediate decuplet, the flavor index should be replaced by its capital counterpart to indicate a range from 1 to 10.}}
\label{fig:boxes}
\end{figure}
Box diagrams including their crossed partners are more involved as the other TME diagrams. The amplitudes of ordinary box diagrams contain two types of poles in the complex {plane} stemming from the baryon and the meson propagators, respectively. The former contribution, however, corresponds just to the first iterate of the Lippmann-Schwinger equation and is thus reducible. The genuine contributions to the TME potential are therefore found by considering the poles of the meson propagators only.

The other thing to note is that a quick view on the diagrams suggests that the potential being $\propto \left(g_{BB\Phi}\right)^4$ seemingly is of $\order{N_c^2}$ which challenges the assumption that the potential should be of $\order{N_c}$. This is exactly the kind of contradiction that has to be remedied by symmetry constraints after including decuplet baryons and combining ordinary box and crossed box diagrams, Figure~\ref{fig:boxes}.

Proceeding in a similar way as for the case of the triangles diagrams, we assume that the resulting potential of both box and crossed box diagrams can be split up into a product of coupling tensors carrying the information on the flavor structure and some function $V^{\Box}_0$
\begin{equation}
V^{\Box/\bowtie}_{B^aB^b\to B^c B^d} \sim  \left(g_{BB\Phi/BT\Phi}\right)^4 V^{\Box}_0\left(|\mathbf{q}|^2, M_{\Phi_i}, M_{\Phi_j}, \lambda\right) .
\end{equation} 
This function $V^{\Box}_0$ is the same for each diagram and both intermediate octet and decuplet baryons to leading order in $1/N_c$ up to some prefactors, as has to be shown. Actually, these yet-to-be-determined prefactors will include a relative minus sign between box and crossed box diagrams that is crucial for the cancellation of the contradictory $N_c^2$ contributions.
\begin{figure*}
\begin{align}
\begin{split}\label{eq:figeq1}
V^{\Box}_0 & = \int \frac{\mathrm d^3k^\prime}{(2\pi)^3}\int\frac{\mathrm dk^\prime_0}{2\pi}\ \frac{\tilde{V}_0^{\Box}(k^\prime,q,M_{\Phi_i}, M_{\Phi_j})}{\left(-k_0^\prime+\frac{|\mathbf{p}|^2}{2m_B}-\frac{|\mathbf{p}-\mathbf{k}^\prime|^2}{2m_B}\right)\left(k_0^\prime+\frac{|\mathbf{p}|^2}{2m_B}-\frac{|\mathbf{p}-\mathbf{k}^\prime|^2}{2m_B}\right)} \left(1 + \order{\frac{1}{N_c}} \right)\\
V^{\bowtie}_0  & = \int \frac{\mathrm d^3k^\prime}{(2\pi)^3}\int\frac{\mathrm dk^\prime_0}{2\pi}\ \frac{\tilde{V}_0^{\bowtie}(k^\prime,q,M_{\Phi_i}, M_{\Phi_j})}{\left(-k_0^\prime+\frac{|\mathbf{p}|^2}{2m_B}-\frac{|\mathbf{p}-\mathbf{k}^\prime|^2}{2m_B}\right)\left(-k_0^\prime-q_0+\frac{|\mathbf{p}|^2}{2m_B}-\frac{|\mathbf{p}+\mathbf{k}^\prime+\mathbf{q}|^2}{2m_B}\right)} \left(1 + \order{\frac{1}{N_c}} \right)
\end{split}
\end{align}\hrule
\end{figure*}

This can be seen when writing down the loop integrals using the notation established in the previous subsection,
\begin{equation*}
\text{see Eq.~\eqref{eq:figeq1} below}
\end{equation*}
where the functions $\tilde{V}_0^{\Box}$ and $\tilde{V}_0^{\bowtie}$ encapsulate the meson propagators that are identical in both cases, and the vertex functions excluding the coupling tensors. Regarding the $k_0^\prime$ integration, we can use the same argument as in the triangle case and substitute the principal values $\mathcal{P}$
\begin{align}
\begin{split}
V^{\Box}_0 & = -i\int \frac{\mathrm d^3k^\prime}{(2\pi)^3} \mathcal{P} \left[ \int\frac{\mathrm dk^\prime_0}{2\pi}\ \frac{\Im\left[\tilde{V}_0^{\Box}(k^\prime,q,M_{\Phi_i}, M_{\Phi_j})\right]}{\left(k_0^\prime\right)^2}\right] ,\\
V^{\bowtie}_0  & = i\int \frac{\mathrm d^3k^\prime}{(2\pi)^3}\mathcal{P} \left[ \int\frac{\mathrm dk^\prime_0}{2\pi}\ \frac{\Im\left[\tilde{V}_0^{\bowtie}(k^\prime,q,M_{\Phi_i}, M_{\Phi_j})\right]}{\left(k_0^\prime\right)^2}\right] , 
\end{split}
\end{align}
giving the relative minus sign mentioned above and a factor of $(1 + \order{1/N_c})$ is implied. Again, without explicitly performing the integrals, we find that $V^{\Box}_0 = - V^{\bowtie}_0$ if $\tilde{V}^{\Box}_0 = \tilde{V}^{\bowtie}_0$. As the meson propagators are the same in both cases, this is just a matter of the vertex functions which involve Pauli matrices in the case of intermediate octet baryons and the spin transition operators given in Eq.~\eqref{eq:spintransitionoperators} in the case of intermediate decuplets. The difference to leading order in $1/N_c$ is just a factor of $2/3$ for each baryon line containing an intermediate decuplet. As in the case of the triangle diagrams, we regard this as a prefactor associated with the coupling tensors, so the total potential stemming from ordinary and crossed box diagrams of both intermediate octet and decuplet baryons is given by
\begin{align}\begin{split}
V^{\Box,\bowtie}_{B^aB^b\to B^c B^d} = & \Biggl(g_{BB\Phi}^{fjb}g_{BB\Phi}^{fdi}-g_{BB\Phi}^{fib}g_{BB\Phi}^{fdj}\\&+\frac{2}{3}\left[\left(g_{BT\Phi}^{Fbj}\right)^\dagger g_{BT\Phi}^{Fdi}-\left(g_{BT\Phi}^{Fbi}\right)^\dagger g_{BT\Phi}^{Fdj}\right]\Biggr)\\
& \times \left(g_{BB\Phi}^{eia}g_{BB\Phi}^{ecj}+\frac{2}{3}\left(g_{BT\Phi}^{Eai}\right)^\dagger g_{BT\Phi}^{Ecj}\right)\\
& \times V^{\Box}_0\left(|\mathbf{q}|^2, M_{\Phi_i}, M_{\Phi_j}, \lambda\right)\\
& \times \left(1 + \order{\frac{1}{N_c}} \right),
\end{split}\end{align}
where implicit sums run over $i,j,e,f=1\dots 8$ and $E,F=1\dots 10$. The full expression of $V_0^\Box$ is presented in the Appendix of Ref.~\cite{Haidenbauer:2013oca} and leads to a central, spin-spin and tensorial part. This leading order expression being seemingly of $\order{N_c^2}$ hence must vanish in order to preserve consistency. This is achieved if the coupling constant $C$ of the baryon-decuplet Lagrangian Eq.~\eqref{eq:decupletlagrangian} takes on the large-$N_c$ value 
\begin{equation}
C = \frac{6}{5}g_A  \left( 1 + \order{\frac{1}{N_c^2}}\right) ,
\end{equation}
which is equivalent to the ratio $C/D = 2$ that is known in the literature \cite{Flores-Mendieta:1999ctw,Flores-Mendieta:2000ljq,Lutz:2001yb}. Note that the correction of $\order{1/N_c^2}$ is neccessary to obtain the overall scaling of $\order{1}$ that is allowed for the two-meson exchange contribution.

\section{Summary}
Starting from the large-$N_c$ baryon-baryon potential derived from a Hartree-like Hamiltonian, we have studied
the large-$N_c$ dependence of the baryon-baryon potential derived from SU(3) chiral perturbation theory assuming
baryon momenta  and strangeness of $\order{1}$. Here, we summarize the results:
\begin{itemize}
\item The baryon-baryon potential is of $\order{N_c}$ and dominated by $V_0^0$, $V_0^1$, $V_\sigma^1$, $V_T^1$,
see Eq.~\eqref{eq:symbolicPot}, corresponding to the central, spin-spin, and tensorial part of the potential.
This is in agreement with the nucleon-nucleon case except for the central part $\sim V_0^1$, which in the
nucleon-nucleon case is of subleading order~\cite{Kaplan:1995yg,Kaplan:1996rk}. The lifting of this term to
$\order{N_c}$ in the $N_f=3$ case is a particularity of the assumption that the large-$N_c$ equivalents of the
real-world nucleons and hyperons are those with strangeness of $\order{1}$ leading to the more complex scalings
of the generator $\hat{\mathcal{T}}_a$ given in Eq.~\eqref{eq:TGscalings} and hence of the term $\sim v_{0,1}^{(T)}$
in the large-$N_c$ potential, Eq.~\eqref{eq:largeNCPot}.
\item The contact terms of leading order in chiral perturbation theory, see Sec.~\ref{sec:leadingordercontact},
generate a potential that includes central, spin-spin, spin-orbit, and tensorial parts. However, only the central
and spin-spin parts $\sim c_S^{abcd}$ and  $\sim c_T^{abcd}$ of this potential are indeed of $\order{N_c}$, while all
other contributions are suppressed by a factor $1/m_B^2$. The contact terms alone hence do not generate the full
leading $\order{N_c}$ potential, but only terms corresponding to $V_0^0$, $V_0^1$, and $V_\sigma^1$ in
Eq.~\eqref{eq:symbolicPot}, while an $\order{N_c}$ tensorial part is missing. Moreover, the spin-orbit part
$\sim c_5^{abcd}$ is of subleading $\order{1/N_c}$ as expected. What these contact terms also add is a partial expansion
of the large-$N_c$ coefficients in Eq.~\eqref{eq:largeNCPot} in the momenta, which can not be determined from the
large-$N_c$ Hartree scenario. All coefficients $c_i^{abcd}$ with $i\neq\{S,T,5\}$ belong to this category.
\item The leading $\order{N_c}$ contact contributions $\sim c_S^{abcd}$ and $\sim c_T^{abcd}$ consist of
linear combinations of six of the original 15 low-energy constants of the contact Lagrangian. In
Section~\ref{sec:largenccontact}, we derived sum rules valid at leading order in $1/N_c$ allowing to reduce the
number of independent parameters to three. Applying these sum rules to the hyperon-nucleon potential studied
in Ref.~\cite{Polinder:2006zh}, see Sec.~\ref{sec:hyperonnucleoncheck}, we were able to use the best-fit values
of the hyperon-nucleon potentials $C_{1S0}^{\Lambda\Lambda}$, $C_{3S1}^{\Lambda\Lambda}$, and $C_{3S1}^{\Lambda\Sigma}$ to
predict the potentials $C_{1S0}^{\Sigma\Sigma}$ and $C_{3S1}^{\Sigma\Sigma}$ finding striking agreement.
\item We have also studied higher order contact terms with explicit insertions of the quark mass matrix.
In general, the resulting potential is structurally similar to the leading order one, but with an extra
suppression of $\order{\epsilon/N_c}$, as these terms involve contributions from SU(3) symmetry breaking of the
order $\epsilon$. Note that for $N_c=3$ the value of $\epsilon/N_c$ has roughly the same magnitude as $1/N_c^2$. However, confronting the hyperon-nucleon potential from chiral perturbation theory with experimental data, such terms can not be neglected, see, e.g., Ref.~\cite{Haidenbauer:2014rna}. 
\item A baryon-baryon potential derived from SU(3) chiral perturbation theory must include one-meson exchange
contributions in order to fully reproduce the leading order large-$N_c$ potential, as the tensorial part $V_T^1$
of $\order{N_c}$ can not be generated by the contact terms alone, which only generate a tensorial part of
$\order{1/N_c}$. This is just in accordance with chiral power counting which also requires the incusion of
leading order contact interactions and one-meson exchanges, see Eq.~\eqref{eq:chiralpower}.
\item Matching the one-meson exchange contributions with the large-$N_c$ potential yields the already known
ratio $F/D = 2/3 \left( 1 + \order{1/N_c^{2}}\right)$, see e.\,g. \cite{Dashen:1993jt}. We also derived an
effective coupling $g_{BB\Phi}$ in terms of $g_A=F+D$ that is valid at leading order in $1/N_c$. In the literature it
is common to use hyperon-nucleon and hyperon-hyperon couplings $f_{BB\Phi}$ expressed in terms of $f_{NN\pi}=g_A/(2F_0)$
and $\alpha=F/(F+D)$ based on Ref.~\cite{deSwart:1963pdg}. The effective large-$N_c$ coupling $g_{BB\Phi}$ just
reproduces these $f_{BB\Phi}$ after forming approriate isospin combinations and setting $\alpha=2/5$.
\item It is also of relevance that the full large-$N_c$ scaling of $\order{N_c}$ in the one-meson exchange
case is only achieved by exchanging pions, while exchanging kaons are of $\order{1}$ and exchanging $\eta$'s are
even more suppressed and of $\order{1/N_c}$ which is a consequence of the choice to match real-world baryons with
those large-$N_c$ baryons that have strangeness of $\order{1}$, see Eq.~\eqref{eq:TGscalings}. At the level of
quarks and gluons, this is just a result of combinatorics, as with this choice there are about $N_c$ choices
to pick up an up or down quark, but only $\order{1}$ choices to find a strange quark. 
\item The large-$N_c$ scalings of many-meson exchange contributions can not be assessed by means of a naive
power counting of the involved meson-baryon couplings alone, as this might lead to results that contradict the
assumption that the baryon-baryon potential is of $\order{N_c}$. However, imposing spin-flavor symmetry and
considering all diagrams of a given type including the full baryon tower retains consistency. Summing over all
$n$-meson exchange diagrams of a given type yields a contribution that at most scales as $\order{N_c^{2-n}}$.
\item For the TME contributions in SU(3) chiral perturbation theory, we showed this explicitly. In this case, the inclusion of
decuplet baryons is mandatory, and a cancellation between the deceptive $\order{N_c^2}$ contributions of the box and
crossed box diagrams appears if the large-$N_c$ ratio $C/D = 2$ in addition to the ratio $F/D= 2/3$. To leading
order it is thus possible to describe one-meson and two-meson exchange diagrams by a single parameter, e.\,g.
by setting $D=3/5g_A$,  $F=2/5g_A$, and $C=6/5g_A$.
\item Among the TME contributions, the box, crossed box, and  triangle diagrams are of $\order{1}$, while
the football diagrams are of subleading $\order{1/N_c^2}$, which is a particularity of chiral perturbation
theory when the number of exchanged mesons is even.
\end{itemize}
The results suggest that a simultaneous expansion in large-$N_c$ and chiral power counting can be used to
reduce the number of ingredients to the baryon-baryon potential at a given order. While it is clear that at
leading order the inclusion of contact interactions $\sim c_S$ and $\sim c_T$ and one-pion exchange diagrams
is obligatory, any extension to higher orders depends on the weight that is assigned to powers $1/N_c$ in
relation to chiral power counting. For instance, one might count powers of $1/N_c \sim \order{q^2}$ as argued
by the authors of Ref.~\cite{Leutwyler:1996sa,Kaiser:2000gs} for the mesonic sector.

However, it seems that such an approach is misleading in the baryonic sector, because some contributions
then appear to be overly suppressed. For instance, in this scheme the SU(3) symmetry breaking contact terms
would count as $\order{q^2/N_c}$ and would show up only way beyond the $1/m_B^2$ corrections of the leading
order terms ($\sim q^0/N_c$) and the box, crossed box, and triangle TME diagrams ($\sim q^2N_c^0$). However,
when confronted with the (still sparse) experimental data of hyperon-nucleon and hyperon-hyperon scattering,
the importance of these SU(3) symmetry breaking contact terms is evident~\cite{Haidenbauer:2014rna}.

The problem here seems to be that such a simultaneous power counting scheme doubly suppresses contributions
that are subleading in terms of both chiral power counting and the $1/N_c$ expansion, even though they are
suppressed for the same reason. This applies, for example, to the $1/m_B^2$ corrections that are treated
as suppression factors in the non-relativistic expansion of chiral perturbation theory relegating such
contributions to higher order, but are also $\order{1/N_c^2}$, which is basically the same statement. Clearly,
this also holds for the SU(3) breaking terms mentioned above, which are of $\order{q^2}$ in terms of chiral
power counting \textit{because} they contain $\mathcal{M}$ and hence signal explicit SU(3) breaking, and are
of $\order{\epsilon/N_c}$ \textit{because} they explicitly break the large-$N_c$ contracted SU(6) symmetry. In a sense,
the power counting of chiral perturbation theory and the large-$N_c$ limit just go hand in hand with each
other regarding these contributions, and what this study shows is that both schemes are mutually consistent.

Consequently, a more cautious approach would be to use the results of the large-$N_c$ analysis to assign
different weights only among the contributions at a given chiral order. So at chiral order $q^0$, the
large-$N_c$ analysis reveals that the contact terms $\sim c_S$ and $\sim c_T$ and one-pion exchange diagrams
are more important than one-kaon exchange diagrams, which in turn are more important than the $1/m_B^2$
contributions and one-$\eta$ exchange diagrams. At chiral order $q^2$, the SU(3) breaking contact terms and
TME box, crossed box, and triangle diagrams are more relevant than the TME football diagram.

\begin{acknowledgement}
This work was supported in part by the European Research Council (ERC) under the
European Union's Horizon 2020 research and innovation programme (grant No.~101018170),
and by the MKW NRW under the funding code NW21-024-A.
The work of UGM was further supported by CAS through the President's
International Fellowship Initiative (PIFI) (Grant No. 2025PD0022).
\end{acknowledgement}

\appendix
{\section{SU(6) commutation relations}\label{app:su6commu}
The 35 generators given in Eq.~\eqref{eq:HartreeGenerators} obey the commutation relations \cite{Dashen:1994qi}
\begin{align}
\com{\hat{\mathcal{S}}^i}{\hat{\mathcal{T}}^a} & = 0, & & \nonumber\\
\com{\hat{\mathcal{S}}^i}{\hat{\mathcal{S}}^j} & = i\epsilon^{ijk} \hat{\mathcal{S}}^k,& \com{\hat{\mathcal{T}}^a}{\hat{\mathcal{T}}^b} & = if^{abc} \hat{\mathcal{T}}^c,\\
\com{\hat{\mathcal{S}}^i}{\hat{\mathcal{G}}^{ja}} & = i\epsilon^{ijk} \hat{\mathcal{G}}^{ka}, &  \com{\hat{\mathcal{T}}_a}{\hat{\mathcal{G}}^{ib}} & =if^{abc} \hat{\mathcal{G}}^{ic} ,\nonumber
\end{align} 
and
\begin{equation}
\com{\hat{\mathcal{G}}^{ia}}{\hat{\mathcal{G}}^{jb}} = \frac{i}{4} \delta^{ij} f^{abc} \hat{\mathcal{T}}^c + \frac{i}{6}\delta^{ab}\epsilon^{ijk}\hat{\mathcal{S}}^k+\frac{i}{2}\epsilon^{ijk}d^{abc}\hat{\mathcal{G}}^{kc}.
\end{equation}}
\section{SU(3) properties and tensor relations}\label{app:su3}
The matching procedure of the previous sections involved manipulations of traces over Gell-Mann matrices and of the two third rank tensors $f$ and $d$ of the respective SU(3) algebra. Throughout this paper, we use the symbols
\begin{align}
\begin{split}\label{eq:htlambda}
h^{abc} & = d^{abc}+if^{abc},\\
t^{abc} & = \frac{1}{2}d^{abc}+\frac{i}{3}f^{abc},\\
\lambda^{a_1a_2\dots a_i} & = \frac{1}{4}\tr{\lambda^{a_1}\lambda^{a_1}\dots\lambda^{a_i}},
\end{split}
\end{align}
which altogether are cyclic in their respective indices. Here, we summerize the most important properties and relations used during our calculations taken from Refs.~\cite{MacFarlane:1968vc,Dittner:1971fy,Borodulin:2017pwh}. The tensors $f$ and $d$ are defined by the commutators and anticommutators of the matrices
\begin{align}
\begin{split}
\com{\lambda^a}{\lambda^b} & = 2if^{abc}\lambda^c,\\
\acom{\lambda^a}{\lambda^b} & = \frac{4}{3}\delta^{ab}\mathbbm{1}+2d^{abc}\lambda^c,\\
\lambda^a\lambda^b &= \frac{2}{3}\delta^{ab}\mathbbm{1}+h^{abc}\lambda^c.
\end{split}
\end{align}
Traces of sequences of Gell-Mann matrices are given by
\begin{equation}
\begin{split}\label{eq:lambdaabcd}
\tr{\lambda^a} & = 0,\\
\tr{\lambda^a\lambda^b} & = 2\delta^{ab},\\
\tr{\lambda^a\lambda^b\lambda^c} & = 2h^{abc},\\
\tr{\lambda^a\lambda^b\lambda^c\lambda^d} & = \frac{4}{3}\delta^{ab}\delta^{cd}+2h^{abk}h^{cdk},\\
\tr{\lambda^a\lambda^b\lambda^c\lambda^d\lambda^e} & = \frac{4}{3}\delta^{ab}h^{cde}+\frac{4}{3}\delta^{de}h^{abc}+2h^{abk}h^{kcl}h^{lde}.
\end{split}
\end{equation}
The tensors $f$ and $d$ obey the Jacobi identities
\begin{align}
\begin{split}
f^{abe}f^{cde} - f^{ace}f^{bde} + f^{bce}f^{ade} & = 0,\\
d^{abe}f^{cde} + d^{ace}f^{bde} + d^{bce}f^{ade} & = 0,
\end{split}
\end{align}
Another useful relation can be found after some algebra
\begin{equation}\label{eq:tacetbde}
6t^{ace}t^{bde} = \frac{25}{12}\lambda^{acbd}+\frac{5}{12}\lambda^{acdb}+\frac{5}{12}\lambda^{cabd}+\frac{1}{12}\lambda^{cadb} -\delta^{ac}\delta^{bd}.
\end{equation}
As it is relevant with respect to the matching procedure, we replicate the non-vanishing values of the SU(3) structure constants (up to permutations):
\begin{align}
\begin{split}\label{eq:appstructureconstants}
f_{123} & = 1,\\
f_{147} = -f_{156} = f_{246} = f_{257} = f_{345} = -f_{367} & = \frac{1}{2}, \\
f_{458} = f_{678} & = \frac{\sqrt{3}}{2}, \\
d_{146} = d_{157}= d_{256} = d_{344} = d_{355}  & = \frac{1}{2}, \\  d_{247} = d_{366} = d_{377} & =- \frac{1}{2}, \\
d_{118} = d_{228} = d_{338} = -d_{888} & = \frac{1}{\sqrt{3}},\\
d_{448} = d_{558} = d_{668} = d_{778} & = -\frac{1}{2\sqrt{3}}.
\end{split}
\end{align}  
For simplifying two-meson exchange contributions, we also used
\begin{equation}\label{eq:appff}
f^{acd}f^{bcd} = 3 \delta^{ab} ,
\end{equation}
and
\begin{align}
\begin{split}\label{eq:appfffdffddfddd}
f^{iaj}f^{jbk}f^{kci} & = -\frac{3}{2} f^{abc},\\
d^{iaj}f^{jbk}f^{kci} & = -\frac{3}{2} d^{abc},\\
d^{iaj}d^{jbk}f^{kci} & = \phantom{-}\frac{5}{6} f^{abc},\\
d^{iaj}d^{jbk}d^{kci} & = -\frac{1}{2} d^{abc}.
\end{split}
\end{align}  
     
\section{Non-relativistic expansion of Dirac tensor matrix elements}\label{app:bbpotential}
Any Dirac field bilinear with any element of the Clifford algebra $\Gamma_i$, Eq.~\eqref{eq:Gammai}, can be rewritten in terms of two-component Pauli spinors $\chi_s$
\begin{equation}\label{eq:GammaToM}
\bar{u}(p_2,s_2)\Gamma_i u(p_1,s_2) = \chi_{s_2}^\dagger M_i(p_2,p_1) \chi_{s_1}^\text{} ,
\end{equation}
where the free positive energy Dirac spinors $u(p,s)$ is given by
\begin{equation}
u(p,s)=\sqrt{\frac{E_p + m}{2m}} \begin{pmatrix}
\chi_s \\ \frac{\boldsymbol{\sigma}\cdot \mathbf{p}}{E_p + m} \chi_s 
\end{pmatrix}
\end{equation}
with $E_p = \sqrt{\mathbf{p}^2 + m^2}$. Expanding the matrix elements for low-energy transfers $q_0$ yields the expressions given in Table~\ref{tab:GammaToM}.
\begin{table*}
\centering
\begin{tabular}{c|c}
$\Gamma_i$ & $M_i(p_2,p_1)$ \\
\hline
$\mathbbm{1}$ & $1 + \frac{\left(\mathbf{p}_1-\mathbf{p}_2\right)^2}{8m^2}+\frac{i}{4m^2}\left(\mathbf{p}_1 \times \mathbf{p}_2\right)\cdot \boldsymbol{\sigma}$\\
$\gamma^0$ & $1 + \frac{\left(\mathbf{p}_1+\mathbf{p}_2\right)^2}{8m^2}-\frac{i}{4m^2}\left(\mathbf{p}_1 \times \mathbf{p}_2\right)\cdot \boldsymbol{\sigma}$ \\
$\gamma^i$ & $\frac{\left(\mathbf{p}_1+\mathbf{p}_2\right)^i}{2m}+\frac{i}{2m}\left(\left(\mathbf{p}_1 - \mathbf{p}_2\right)\times \boldsymbol{\sigma}\right)^i$ \\
$\sigma^{0j}$ & $i\frac{\left(\mathbf{p}_1-\mathbf{p}_2\right)^i}{2m}-\frac{1}{2m}\left(\left(\mathbf{p}_1 + \mathbf{p}_2\right)\times \boldsymbol{\sigma}\right)^i$\\
$\sigma^{ij}$ & $\left\{\left(1 + \frac{\left(\mathbf{p}_1+\mathbf{p}_2\right)^2}{8m^2}\right) \sigma^k - \frac{1}{4m^2}\left[i\left(\mathbf{p}_1 \times \mathbf{p}_2\right)^k+p_1^k\sdot{\mathbf{p}_2}{\boldsymbol{\sigma}}+p_2^k\sdot{\mathbf{p}_1}{\boldsymbol{\sigma}}\right]\right\}\epsilon_{ijk}$ \\
$\gamma^i\gamma_5$ & $\left(1 + \frac{\left(\mathbf{p}_1-\mathbf{p}_2\right)^2}{8m^2}\right) \sigma^i + \frac{1}{4m^2}\left[i\left(\mathbf{p}_1 \times \mathbf{p}_2\right)^i+p_1^i\sdot{\mathbf{p}_2}{\boldsymbol{\sigma}}+p_2^i\sdot{\mathbf{p}_1}{\boldsymbol{\sigma}}\right]$ \\
$\gamma^0\gamma_5$ & $\frac{1}{2m} \left(\mathbf{p}_1+\mathbf{p}_2\right)\cdot\boldsymbol{\sigma} + \frac{q_0}{8m^2} \left(\mathbf{p}_1-\mathbf{p}_2\right)\cdot\boldsymbol{\sigma}$\\
$\gamma^5$ & $\frac{1}{2m} \left(\mathbf{p}_1-\mathbf{p}_2\right)\cdot\boldsymbol{\sigma} + \frac{q_0}{8m^2} \left(\mathbf{p}_1+\mathbf{p}_2\right)\cdot\boldsymbol{\sigma}$
\end{tabular}
\caption{Equivalence of $\Gamma_i$ and $M_i(p_2,p_1)$ as defined in Eq.~\eqref{eq:GammaToM}.}\label{tab:GammaToM}
\end{table*}

\end{document}